\documentclass[lettersize,journal]{IEEEtran}
\usepackage{amsmath,amsfonts}
\usepackage{algorithmic}
\usepackage{algorithm}
\usepackage{array}
\usepackage[caption=false,font=normalsize,labelfont=sf,textfont=sf]{subfig}
\usepackage{textcomp}
\usepackage{stfloats}
\usepackage{url}
\usepackage{verbatim}
\usepackage{graphicx}
\usepackage[nocompress]{cite}
\usepackage{threeparttable}
\usepackage{booktabs}
\usepackage{multirow}
\usepackage{tabularray}
\usepackage{amssymb}
\usepackage{pifont}  
\usepackage[dvipsnames]{xcolor}
\usepackage{hyperref}
\hypersetup{
    colorlinks=true,
    linkcolor=Maroon,
    anchorcolor=blue,
    citecolor=RoyalBlue
}

\newcommand{\cmark}{\ding{51}}   
\newcommand{\xmark}{\ding{55}}   
\begin{document}

\title{SurfAge-Net: A Hierarchical Surface-Based Network for Interpretable Fine-Grained Brain Age Prediction}

\author{Rongzhao He,
        Dalin Zhu,
        Ying Wang,
        Songhong Yue,
        Leilei Zhao,
        Yu Fu,
        Dan Wu,
        Bin Hu,~\IEEEmembership{Fellow,~IEEE},
        Weihao Zheng,~\IEEEmembership{Member,~IEEE}
\thanks{This work was supported in part by the National Natural Science Foundation of China under Grant 62227807 and 62202212, in part by the Gansu Provincial Health Commission Scientific Research Program of China under Grant GSWSKY2023-35.(\textit{Rongzhao He and Dalin Zhu contribute equally to this work.}) (\textit{Corresponding author: Weihao Zheng; Dan Wu; Bin Hu.})}
\thanks{This work involved human subjects or animals in its research. Approval of all ethical and experimental procedures and protocols was granted by the Ethics Committee of Gansu Provincial Maternity and Child-Care Hospital under Application No. 2020-GSFY-05 and Lanzhou University Second Hospital under Application No. 2021A-601.}
\thanks{Rongzhao He, Ying Wang, Yu Fu, and Weihao Zheng are with the Gansu Provincial Key Laboratory of Wearable Computing, School of Information Science and Engineering, Lanzhou University, Lanzhou 730000, China (e-mail: herongzhao23@lzu.edu.cn; wangying22@lzu.edu.cn; fuyu@lzu.edu.cn; zhengweihao@lzu.edu.cn).}
\thanks{Dalin Zhu is with the Department of Medical Imaging Center, Gansu Provincial Maternity and Child-Care Hospital, Lanzhou 730050, China (e-mail: zdlldz@126.com).}
\thanks{Songhong Yue is with the Department of Magnetic Resonance, Lanzhou University Second Hospital, Lanzhou 730030, China (e-mail: yshwjt@163.com).}
\thanks{Leilei Zhao is with the Department of Computer Science and Technology, Harbin Institute of Technology (Shenzhen), Shenzhen 518055, China (e-mail: 24B951025@stu.hit.edu.cn).}
\thanks{Dan Wu is with the Key Laboratory for Biomedical Engineering of Ministry of Education, Department of Biomedical Engineering, College of Biomedical Engineering \& Instrument Science, Zhejiang University, Hangzhou 310027, China (e-mail: danwu.bme@zju.edu.cn).}
\thanks{Bin Hu is with the Gansu Provincial Key Laboratory of Wearable Computing, School of Information Science and Engineering, Lanzhou University, Lanzhou 730000, China, also with the School of Medical Technology, Beijing Institute of Technology, Beijing 100081, China (e-mail: bh@lzu.edu.cn).}}

\maketitle

\begin{abstract}
Brain age prediction serves as a powerful framework for assessing 
brain status and detecting deviations associated with 
neurodevelopmental and neurodegenerative disorders. However, most 
existing approaches emphasize whole-brain age prediction and therefore 
overlook the pronounced regional heterogeneity of brain maturation that 
is crucial for detecting localized atypical trajectories. To address this 
limitation, we propose a novel spherical surface-based brain age 
prediction network (SurfAge-Net) that leverages multiple morphological 
metrics to capture region-specific developmental patterns with enhanced 
robustness and clinical interpretability. SurfAge-Net establishes a new 
modeling paradigm by incorporating the connectomic principles of cortical 
organization: it explicitly models both intra- and inter-hemispheric 
dependencies through a spatial-channel mixing and a lateralization-aware 
attention mechanism, enabling the network to characterize the coordinate 
maturation pattern uniquely associated with each target region. Validated 
on three fetal and neonatal datasets, SurfAge-Net outperforms existing 
approaches (global MAE = 0.54, regional MAE = 0.45 in gestational/postmenstrual weeks) and 
demonstrates strong generalizability across external cohorts. Importantly, 
it provides spatially precise and biologically interpretable maps of 
cortical maturation, effectively identifying heterogeneous delays and 
regional-specific abnormalities in atypical developmental populations. 
These results established fine-grained brain age prediction as a 
promising paradigm for advancing neurodevelopmental research and 
supporting early clinical assessment.
\end{abstract}

\begin{IEEEkeywords}
Cortical spherical surface, Fine-grained, Brain age prediction, Fetal and neonatal, Morphological metrics, Interpretability.
\end{IEEEkeywords}

\section{Introduction}
\IEEEPARstart{T}{he} human brain undergoes a complex and dynamic, and highly organized developmental 
process\cite{dehaene2015infancy,wilson2021development}. 
Among the biological indicator used to characterize this trajectory, 
brain age has emerged as a valuable marker reflecting the structural 
and functional maturity of the brain\cite{franke2012brain,whitmore2023brainage}. 
By comparing predicted brain age with chronological age, researchers 
can quantify deviations from normative developmental patterns, 
which are often associated with neurodevelopmental and 
neurodegenerative conditions\cite{licht2009brain,hazlett2017early}. 
For example, accelerated brain development and aging have been 
linked to autism spectrum disorder\cite{hazlett2017early} and 
Alzheimer's disease\cite{he2021global}, respectively; whereas delayed brain 
maturation is commonly observed in preterm infants\cite{zheng2023preterm,dimitrova2021preterm}, 
which provides only a global summary of brain maturity and fails to account 
for substantial regional diversity of developmental trajectories\cite{wang2024profiling,shaw2008neurodevelopmental,huang2022mapping,zheng2023spatiotemporal}. 
As a result, subtle but clinically meaningful regional deviations may be masked when aggregated across the entire brain.

Accumulating evidence indicates that brain maturation progresses in a 
markedly regionally heterogeneous manner. Under typical development, 
cortical maturation follows a unimodal-to-transmodal gradient, with 
primary cortices (e.g., primary sensory and motor regions) and their 
associated white-matter tracts mature earlier during late gestation and 
early infancy, while higher-order association regions (e.g., default mode and salience networks) 
mature substantially later\cite{dehaene2015infancy,zheng2023spatiotemporal,thomason2020development}. 
However, this maturational gradient is often disrupted in atypical 
developmental populations. For example, preterm infants show selectively 
delayed neurite maturation in posterior cortical regions but accelerated 
development of cortical thickness in anterior cortex compared to term 
newborns\cite{dimitrova2021preterm}. Similarly, abnormal expansion of prefrontal surface area has 
been observed in high-risk infants who were diagnosed with autism at 24 
months\cite{hazlett2017early}. Such regional variability during atypical development may lead 
to region-specific deviations in developmental status that cannot be 
adequately captured by a global brain-age model, thereby underscoring the 
need for spatially resolved models capable of quantifying localized cerebral maturity.

Nevertheless, despite this widely demonstrated regional developmental 
heterogeneity, most existing approaches remain confined to estimating a 
single global brain-age value\cite{wu2024comparative}. While such a global metric provides a 
concise summary of overall brain maturation, it inevitably collapses the 
diverse regional developmental trajectories distributed across the cortex. 
This spatial averaging may obscure clinical meaningful deviation, e.g., 
accelerated maturation in anterior cortex of preterm infants, that are 
closely associated with subsequent cognitive development, functional 
specialization, and domain-specific behavioral outcomes. In addition, 
previous studies have predominantly emphasized model interpretability, 
typically achieved through visualizing feature contribution or attention 
weights for the prediction\cite{dahan2022surface,li2025surfgnn}. However, this form of interpretability 
does not necessarily translate into clinical interpretability, which 
concerns \textit{where}, \textit{in what direction}, and \textit{to what extent} an individual's brain 
development deviates from normative trajectories, as well as the potential 
cognitive or behavioral implications of such deviations. Therefore, the 
clinical utility and risk-assessment relevance of global brain age remain 
inherently limited, motivating a growing interest in fine-grained brain-age 
estimation. For example, several studies have attempted voxel-level brain-age 
prediction\cite{cole2017predicting,dartora2024deep}. While these approaches markedly enhance spatial resolution 
relative to whole-brain estimates, the extremely high dimensionality of 
voxel-wise representations significantly increases the risk of overfitting 
and computational burden. Moreover, voxel-level predictions are highly 
sensitive to noise or preprocessing variability (e.g., registration inaccuracies and spatial smoothing effects), 
which further weakened their robustness. These limitations indicate that 
finer granularity is not inherently advantageous. Instead, it is essential 
to define an appropriate representational scale that balances predictive 
precision, spatial resolution, computational efficiency, and model robustness.

Surface-based morphometry (SBM) has become a powerful tool for characterizing cortical 
morphology. By mapping the highly convoluted cortex onto a spherical surface, SBM enables 
more accurate quantification of complex geometric structures and folding patterns. Compared 
to voxel-based morphometry (VBM), SBM offers improved cortical alignment and parcellation, 
which reduces errors introduced by complex gyral-sulcal topography, and provides a richer 
set of morphological descriptors, such as cortical thickness, sulcal depth, and curvature, 
that jointly capture complementary aspects of cortical shape\cite{spalletta2018brain}. Recent studies have 
demonstrated strong age-related changes in these features, each exhibiting distinct 
developmental or aging trajectories across the lifespan\cite{wang2023age,bethlehem2022brain}. Such feature-specific 
heterogeneity is especially pronounced during the first year of life, a period marked by 
rapid and highly nonuniform cortical growth\cite{xu2022spatiotemporal,sheng2024no}. Given that SBM depicts multiple 
dimensions of cortical morphology that are often missing in voxel-based representation, 
it may provide a more comprehensive and biologically sensitive basis for brain age estimation. 
In fact, recent studies have shown the effectiveness of employing multiple SBM metrics for 
this purpose, particularly in the neonatal cohort\cite{he2025surface,zhao2024attention,dahan2022surface}, 
supporting the potential of SBM to enable more precise characterization of brain 
maturational status, especially when the goal is to capture subtle, spatially localized developmental patterns.

\begin{figure}
	\includegraphics[width=1\columnwidth]{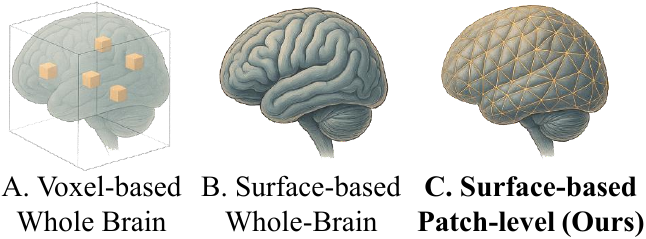}
	\caption{Different brain age prediction paradigms.}
    \label{fig1}
\end{figure}

In the present study, we propose a patch-level brain age prediction 
paradigm based on cortical surface data, termed SurfAge-Net. An overview of different brain 
age prediction paradigms is illustrated in \hyperref[fig1]{Fig. \ref*{fig1}}. The primary 
aim of SurfAge-Net is to enhance the clinical interpretability of brain-age estimates by 
providing a spatially resolved mapping of maturational heterogeneity throughout the cortex, 
meanwhile maintaining satisfied prediction accuracy. To achieve this goal, we introduce a 
connectome-inspired modeling strategy that allows the age of each cortical patch to be 
predicted not only from its own morphological metrics but also from its intrinsic 
interactions with anatomically related regions. Specifically, we design a Spatial-Channel 
Mixing Block to facilitate effective information exchange among intra-hemispheric 
neighboring patches, and incorporate a Lateralization-Aware Attention Mechanism to 
selectively integrate contralateral information. In addition, a Gated Filter Module is 
introduced, which adaptively balance ipsilateral and contralateral contributions while 
suppressing irrelevant patches, reflecting the brain connectivity principles that cortical 
areas typically interact with a subset of structurally or functionally meaningful 
partners\cite{zheng2019multi,sporns2004organization}. These components enable SurfAge-Net to learn biologically informed 
inter-regional dependencies that are essential for precise and clinically meaningful 
patch-level brain-age estimation. The main contributions of this work are summarized as follows:
\begin{enumerate}
  \item We introduce SurfAge-Net, the first spherical-surface-based patch-level prediction framework for fine-grained brain-age estimation, enabling high-resolution assessment of cerebral regional maturity.
  \item SurfAge-Net achieves state-of-the-art performance in both global and regional brain-age prediction and demonstrates strong generalizability across multiple external cohorts.
  \item SurfAge-Net substantially enhances clinical interpretability by generating spatially resolved cortical age maps that uncover localized maturational deviations in atypically developing populations.
\end{enumerate}

\section{Related Work}

\subsection{Global Brain Age Prediction}
Early research on global brain-age prediction primarily relied on statistical learning models 
with handcrafted features\cite{chung2018use,becker2018gaussian,hu2019hierarchical,beheshti2022accuracy,liu2022brain}. 
These works demonstrated the feasibility of using brain age as a biomarker for neurodevelopmental 
and neurodegenerative conditions, yet their predictive performance remained limited due to 
shallow learning ability and the restricted expressiveness of manually engineered features. 
With the rise of deep learning, convolutional neural networks (CNNs) and their variants\cite{cole2017predicting,huang2017age,ueda2019age,jonsson2019brain,feng2020estimating,bashyam2020mri} 
enabled end-to-end modeling of volumetric MRI, substantially improving performance. 
More recently, geometric deep learning (GDL) has extended brain-age predictions to 
non-Euclidean domains, including cortical surfaces and brain connectivity 
graphs\cite{li2025surfgnn,fawaz2021benchmarking,vosylius2020geometric,monti2017geometric,cohen2018spherical,defferrard2016convolutional,kipf2016semi,zhao2019spherical,qi2017pointnet++}, 
enabling more faithful representation of the intrinsic 
topology of brain. Attention-based architectures have further strengthened the modeling of long-range 
dependencies\cite{dahan2022surface,zhao2024attention,zhao2024transformer,dahan2024multiscale}, 
and the emerging state space models (e.g. Mamba)\cite{gu2024mamba} have provided a favorable trade-off between 
predictive performance and computational efficiency. Despite this methodological progress, 
global prediction frameworks collapse the complex spatial heterogeneity of neurodevelopmental 
into a single scalar measure. Therefore, they are inherently limited in capturing localized 
maturational deviations that are critical for understanding heterogeneous developmental 
process and disease-specific brain alterations.

\begin{table*}
\centering
\begin{threeparttable}
    \caption{Demographic information of participants used in our study.}
    \label{tab1}
    
    \begin{tabular}{l c c c}
    \hline
        ~ & \textbf{\textit{dHCP}} & \textbf{\textit{GSMCH}} & \textbf{\textit{LZUSH}} \\ 
        ~ & \textbf{\textit{(N=1166)}} & \textbf{\textit{(N=65)}} & \textbf{\textit{(N=49)}} \\ \hline 
        \textbf{Birth age [weeks$^{\textbf{+days}}$], median (IQR)} & 37$^{+2}$ (34$^{+6}$-40$^{+4}$) & 33$^{+3}$ (30$^{+4}$-36$^{+4}$) & - \\ 
        \textbf{Scan age [weeks$^{\textbf{+days}}$], median (IQR)}  & 37$^{+4}$ (34$^{+0}$-42$^{+0}$) & 38$^{+3}$ (36$^{+1}$-40$^{+2}$) & 29$^{+6}$ (28$^{+0}$-32$^{+0}$) \\ 
        \textbf{Birth weight$^{\textbf{1}}$, mean (SD)} & 2.83 (0.96) & 1.95 (0.86) & - \\ 
        \textbf{Head circumference at scan$^{\textbf{2}}$, mean (SD)} & 34.16 (2.81) & 32.90 (3.28) & - \\ 
        \textbf{Radiology score (1/2/3/4/5)} & 716/257/122/55/16 & - & - \\ 
        \textbf{Gender$^{\textbf{3}}$ (M/F)} & 631/533 & 34/31 & -/- \\ \hline

    \end{tabular}
    \begin{tablenotes}[flushleft]
        \item $^{1}$ 1 and 2 birth weight data was missed in \textit{dHCP} and \textit{GSMCH}, respectively. 
        \item $^{2}$ 34 and 37 head circumference data were missed in \textit{dHCP} and \textit{GSMCH}, respectively.
        \item $^{3}$ Gender information was unavailable for 2 fetuses in \textit{dHCP}.
	  \end{tablenotes}
\end{threeparttable}
\end{table*}

\subsection{Fine-Grained Brain Age Prediction}
Early efforts towards fine-grained brain-age estimation largely relied on linear 
regression\cite{cherubini2016importance}, most notably voxel-wise prediction through fitting independent linear 
models\cite{kaufmann2019common}. However, such approaches ignored contextual dependencies and fail to capture 
the inter-regional structural coherence that underpin cortical development. Moreover, 
accumulating evidence shows that brain maturation follows regionally distinct and often 
non-linear trajectories\cite{hof1996neuropathological,raz2010trajectories}, 
exceeding the representational capacity of linear models. 
The introduction of deep learning brought significant progress. A U-Net-based voxel-wise 
framework\cite{popescu2021local} generated spatially resolved brain age maps that revealed disease-related 
patterns, such as elevated age gaps in hippocampal and basal ganglia regions in individuals 
with mild cognitive impairment or dementia. More recently, Gianchandani \textit{et al}.\cite{gianchandani2023voxel} 
extended this line of work with a multi-task U-Net that jointly predicted both voxel-level 
and global brain age while performing tissue segmentation, thereby enhancing interpretability. 
Despite these advances, voxel-level approaches face several fundamental limitations. First, 
voxel-based morphometry primarily captures volumetric information and struggles to 
represent cortical geometry, such as folding patterns, which are more naturally captured 
on the cortical surface. Second, voxel-wise representations are extremely high dimensional, 
requiring large training datasets and substantial computational resources, and may also 
increase the risk of overfitting and reduce the robustness to noise and motion artifacts. 
These challenges highlight the need for a method that adopt an appropriate scale and 
meaningful feature representation to enable more reliable fine-grained brain-age prediction.

\begin{figure}  
  \centering
	\includegraphics[width=1\columnwidth]{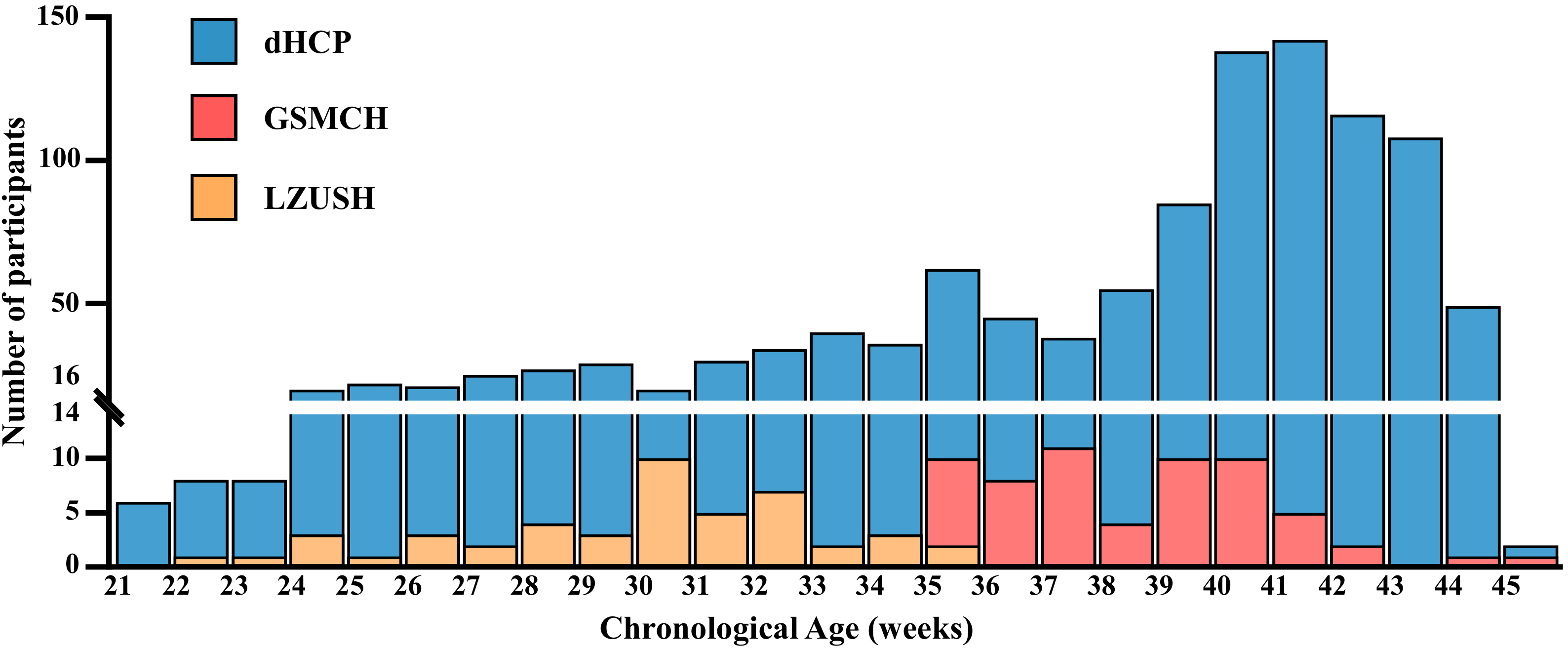}
	\caption{The distribution of chronological ages for all participants utilized in this study.}
    \label{fig2}
\end{figure}

\section{Materials and Methods}

\subsection{Imaging Data}

Three datasets, with a total of 1280 fetal and neonatal MRIs, were used 
in this study: (\textit{i}) the fourth data release of the publicly available 
Developing Human Connectome Project (dHCP), which includes 1166 MRI 
scans of both neonates and fetuses; (\textit{ii}) 65 neonatal brain MRI data 
collected from the Gansu Provincial Maternity and Child-Care 
Hospital (GSMCH); and (\textit{iii}) 49 fetal brain MRI data collected from the 
Lanzhou University Second Hospital (LZUSH). The demographic information 
of the three cohorts is summarized in \hyperref[tab1]{Table \ref*{tab1}}, 
and the distribution of chronological ages for all participants utilized in this study 
is visualized in \hyperref[fig2]{Fig. \ref*{fig2}}.

Images for the \textit{dHCP} dataset were acquired on a 3.0 T Philips Achieva 
scanner equipped with a 32-channel neonatal head coil. For the \textit{GSMCH} 
dataset (approved by ethics committee of \textit{GSMCH}, No. 2020-GSFY-05) were 
acquired using a Siemens Magnetom Lumina 3.0 T scanner with a spin 
echo (SE) sequence. The imaging resolution for both datasets was 
0.8$\times$0.8$\times$1.6 mm$^{3}$ with 0.8 mm overlap, and all images 
were reconstructed to an isotropic resolution of 0.5 mm. For the \textit{LZUSH} 
dataset (approved by ethics committee of \textit{LZUSH}, No. 2021A-601), 
T2-weighted (T2w) fetal brain MRI data were acquired using Single-shot 
Fast Spin Echo (SSFSE) sequence on a Philips Ingenia 3.0 T scanner. 
To mitigate motion artifacts, each fetal brain was scanned multiple 
times in three orthogonal planes (axial, coronal and sagittal), with 
a repetition time of 12,000 ms, an echo time of 80 ms, a matrix size of 
236 $\times$ 220, a flip angle of 90°, a field of view of 260 $\times$ 355 mm², 
slice thickness of 2 mm with no gap, and a scan time ranging from 15 to 45 s. 
As the data were obtained as thick-slice 2D stacks, super-resolution 
reconstruction was required to generate a high-resolution 3D volume, 
and the detailed procedure is described in \hyperref[3.2]{Section \ref*{3.2}}.

To construct a normative model of typical brain development during the 
second and third trimesters, we selected fetal and neonatal images from 
the \textit{dHCP} dataset based on the following criteria: (\textit{i}) only the first 
scan of each fetus/neonate was retained; (\textit{ii}) term-born neonates 
scanned within one week after birth, with no focal 
abnormalities (radiology score $<$ 4). A total of 976 images were 
obtained and divided into training, validation, and test sets with a 
ratio of 8:1:1 within each label interval for training in each infant 
and fetal group. Subsequently, the best-performed model was applied to 
several atypical developmental groups to investigate brain developmental 
characteristics under different atypical conditions, including: (\textit{i}) 66 
and 24 moderate-to-late preterm (32-37 weeks) from \textit{dHCP} and \textit{GSMCH}, 
respectively; (\textit{ii}) 53 and 25 extremely preterm (24-32 weeks) from \textit{dHCP} 
and \textit{GSMCH}, respectively; (\textit{iii}) 71 focal 
abnormalities (radiology score $>$ 3) from \textit{dHCP}.

\begin{figure*}
	\includegraphics[width=\textwidth]{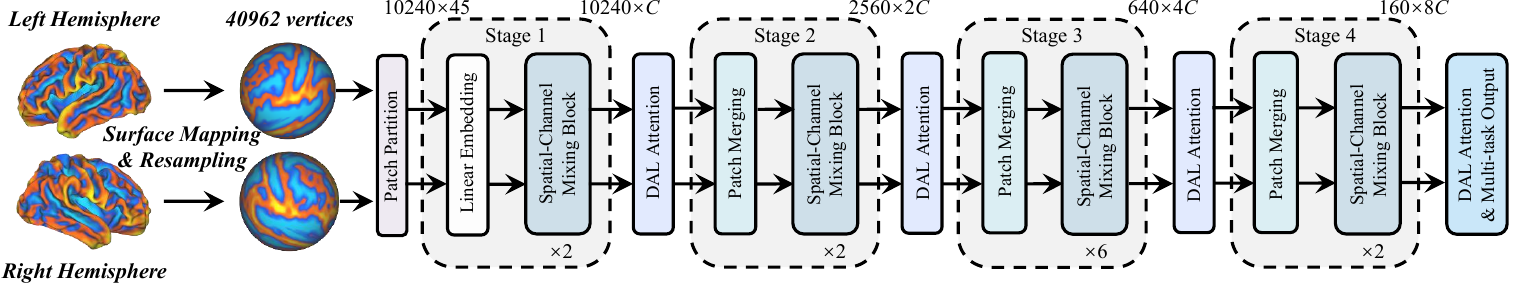}
	\caption{Overview of the SurfAge-Net architecture.}
    \label{fig3}
\end{figure*}

\begin{figure}
	\includegraphics[width=\columnwidth]{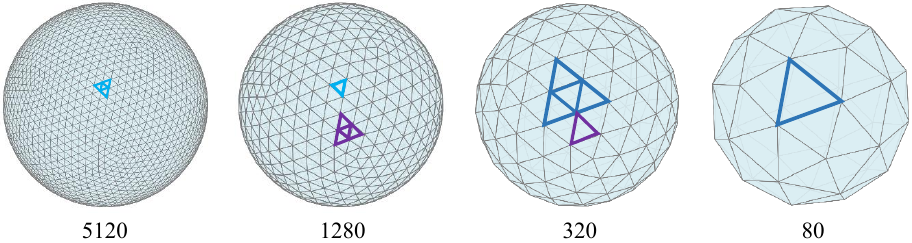}
	\caption{Patch merging process. The number of faces of each spherical surface is denoted under the surface.}
    \label{fig4}
\end{figure}

\subsection{Data Preprocessing}
\label{3.2}
Motion-corrected T2w images underwent cortical reconstruction using the dHCP neonatal structural 
pipeline\cite{makropoulos2018developing}. This framework integrates probabilistic tissue segmentation via 
Draw-EM\cite{makropoulos2014automatic} with surface topology derivation and refinement\cite{schuh2017deformable}, 
followed by generation of anatomically consistent inner (white matter), outer (pial), 
midthickness, inflated and spherical surface representations with vertex-wise correspondence 
across subjects. In addition, the pipeline yields multiple cortical surface phenotypes 
characterising morphology and folding architecture, including cortical thickness, 
sulcal depth and mean curvature. A full methodological description of image reconstruction and 
neonatal-specific preprocessing is detailed in Makropoulos \textit{et al}.\cite{makropoulos2018developing}.

For the \textit{LZUSH} dataset, super-resolution reconstruction was performed to generate 
high-resolution 3D fetal brain volumes. The procedure was as follows. First, fetal 
brain masks were obtained using the NiftyMIC pipeline (\url{https://github.com/gift-surg/NiftyMIC}) 
based on multi-planar 2D stacks (axial, coronal and sagittal). Next, the masked 
stacks were used for super-resolution reconstruction, producing a 3D fetal brain 
volume with 0.8 mm isotropic resolution. The intracranial region was then extracted 
using the reconstructed brain mask, followed by manual AC-PC alignment of the 3D 
MRI volume in MATLAB using SPM.

All data were registered to the dHCP 36-week spherical template, which represents 
the cortical surface as an approximated sphere composed of triangles, with 32,492 
vertices per hemisphere. We resampled the template sphere to a regular sixth-order 
icosphere (Ico-6) using barycentric interpolation. Morphological metrics, including 
mean curvature, sulcal depth, and thickness of cerebral cortex, were calculated 
from T2w images. Each feature channel was normalized using Z-score.

\subsection{Network Architecture}
As shown in \hyperref[fig3]{Fig. \ref*{fig3}}, SurfAge-Net incorporates a novel and well-suited Spatial-Channel 
mixing block (SCM) and Dynamic Adaptive Lateralization-Aware Attention (DALA), 
dedicated to intra-hemispheric and inter-hemispheric modeling, respectively. 
The SCM block enables information exchange among cortical regions within each 
hemisphere while simultaneously enhancing the feature representation of each region. 
In contrast, the DALA module models cross-hemispheric interactions between the 
left and right hemispheres, effectively characterizing the lateralized 
developmental structure of brain and improving prediction performance.

To be more specific, the cortical surface was partitioned into $2N$ patches, denoted as 
$\widetilde{X} = \{\widetilde{L}, \widetilde{R}|\widetilde{L}\in\mathbb{R}^{N \times V \times C}, \widetilde{R}\in\mathbb{R}^{N \times V \times C}\}$, 
where $V$ is the number of vertices per patch and $C$ is the number 
of feature channels. Each patch was then flattened into 
$X = \{L, R|L\in\mathbb{R}^{N\times(VC)}, R\in\mathbb{R}^{N\times(VC)}\}$, 
with $L=[X_L^1,X_L^2, \cdots ,X_L^N ]$ and $R=[X_R^1,X_R^2, \cdots ,X_R^N ]$. 
These representations were processed by a four-stage hierarchical network. 
The first stage applied a linear embedding to project inputs into feature space, 
while the remaining three stages began with a patch-merging layer for progressive 
spatial reduction (shown in \hyperref[fig4]{Fig. \ref*{fig4}}). 
The input data were first partitioned into 10,240 non-overlapping triangular 
patches using an Ico-4 spherical grid. After patch-merging operations in the last 
three stages, the network aggregated these into 160 cortical regions comprising 80 
regions per hemisphere. Each stage consisted of a SCM block and a DALA bock. 
The SCM block integrates information across patches within the same hemisphere to 
capture intra-hemispheric dependencies, and the DALA block models inter-hemispheric 
interactions to capture long-range structural associations in a lateralization-aware manner. 
Finally, SurfAge-Net employed a multi-task prediction head. Each of the 160 cortical 
regions was assigned an independent linear layer to estimate its regional brain age, 
while an additional linear layer was dedicated to whole-brain age prediction. 
This design encourages shared representations that jointly benefit regional and 
global predictions, improving robustness against local noise and interpretability 
in developmental assessment.

\begin{figure*}  
  \centering
	\includegraphics[width=0.8\textwidth]{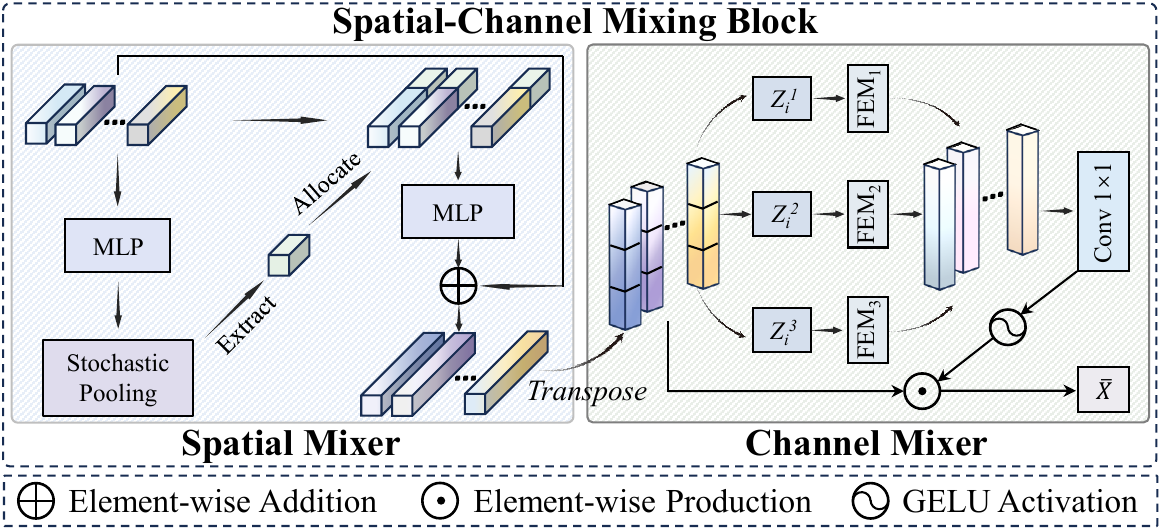}
	\caption{The detailed structures of the Spatial-Channel Mixing Block.}
    \label{fig5}
\end{figure*}

\subsection{Spatial-Channel Mixing Block}
The human brain exhibits strong structural lateralization, 
with hemisphere-specific morphometric patterns in cortical thickness, 
sulcal depth, and curvature. These patterns are not uniformly distributed but 
instead follow hemisphere-specific folding architectures and developmental 
constraints. To preserve these signatures, we first performed intra-hemispheric 
modeling by aggregating hemisphere-level information and enhancing patch-wise 
features for regional brain age prediction. An MLP-based design was employed for 
spatial-channel mixing, achieving linear complexity with respect to the number of 
patches and avoiding the computational burden of global attention. The detailed 
structure of SCM is shown in \hyperref[fig5]{Fig. \ref*{fig5}}.

\subsubsection{Spatial Mixer}
The Spatial Mixer is designed as a resource-efficient module for global 
information aggregation within each hemisphere. Given the patch-wise feature 
representation $X$, we first projected the input into $D$-dimensional vectors 
using a parameterized MLP. To obtain a hemisphere-level summary, we applied 
stochastic pooling across the $N$ patches. For each feature dimension $j$, the 
selection probability of patch $i$ is defined by softmax normalization:
\begin{equation}
  p_{ij}=\frac{e^{x_{ij}}}{\sum_{k=1}^{N}e^{x_{ij}}}
\end{equation}
where, $x$ is $L$ or $R$; stochasticity is introduced by sampling one patch index $c$ 
from the Categorical distribution parameterized by the probabilities during training:
\begin{equation}
  o_{j}=x_{cj},\quad c\sim Categorical(p_{1j,}p_{2j,\cdots,}p_{Nj})
\end{equation}
where “Categorical” denotes a discrete probability distribution 
parameterized by the vector $(p_{1j,}p_{2j,\cdots,}p_{Nj})$. 
The stochastic sampling introduces randomness during training, acting as a form of 
regularization that encourages robust representations learning. 
At inference time, however, stochastic sampling would lead to non-deterministic 
outputs, therefore we replaced it with its mathematical expectation to ensure 
consistency and reproducibility:
\begin{equation}
  o_j=\sum_{i=1}^Np_{ij}x_{ij}
\end{equation}
The pooled outputs across all feature dimensions were concatenated to form the Global Representation (GR):
\begin{equation}
  GR=[o_1,o_2,\cdotp\cdotp\cdotp,o_{D^{\prime}}]\in\mathbb{R}^{1\times D^{\prime}}
\end{equation}
To enhance information interactions within each hemisphere, the GR was concatenated with each patch features:
\begin{equation}
  \widetilde{x_{i}}=MLP([x_{i}\|GR])
\end{equation}
where $||$ denotes concatenation. Finally, residual connections were applied to avoid gradient diminishing:
\begin{equation}
  {x_{i}}^{\prime}=x_{i}+\widetilde{x_{i}}
\end{equation}
\subsubsection{Channel Mixer}
Following spatial aggregation, we employed a Channel Mixer to enhance 
representational power at the patch level. Inspired by the multi-head attention 
mechanism, we split each patch feature vector into three disjoint channel groups, 
enabling complementary specialization across channels and improving the encoding 
of fine-grained details. Formally, we transposed the outputs to reorganize features 
along the feature dimension, yielding the representation:
\begin{equation}
  Z=[z_1,z_2,\cdots,z_N],\quad z_i\in\mathbb{R}^{1\times D}
\end{equation}
where $z_i$ denotes the feature of the $i$-th patch. To further enrich patch-level 
representations, we uniformly divided the feature channels of each patch into three 
disjoint subspaces, which were processed through a Feature Enhancement Module (FEM):
\begin{equation}z_i=
\begin{bmatrix}
z_i^{c_1},z_i^{c_2},z_i^{c_3}
\end{bmatrix},\quad z_i^{c_m}\in\mathbb{R}^{1\times\frac{D}{3}},\quad m=1,2,3
\end{equation}
For each subspace, we apply an independent transformation function 
$f_{m}(\cdot)$. The first group was processed directly using a depthwise convolution:
\begin{equation}
  f_1\left(z_i^{c_1}\right)=DWConv_1\left(z_i^{c_1}\right)
\end{equation}
The remaining two groups were processed in a multi-scale manner. Each subset was 
adaptively pooled to a reduced dimension $p_m$, and then upsampled back to the 
original feature length via nearest-neighbor interpolation:
\begin{equation}
  f_m(z_i^{c_m})=Up_{\uparrow\frac{D}{3}}\left(DWConv\left(Pool_{\downarrow p_m}(z_i^{c_m})\right)\right),m=2,3
\end{equation}
where $Pool_{\downarrow p_m}(\cdot)$ represents adaptive max pooling and $Up_{\uparrow\frac{D}{3}}(\cdot)$
restores the channel dimension to $\frac{D}{3}$.

The transformed groups were concatenated and aggregated to form the enhanced patch representation:
\begin{equation}
  h_i=Conv([f_1(z_i^{c_1})\|f_2(z_i^{c_2})\|f_3(z_i^3)])
\end{equation}
Finally, a nonlinear activation was applied to the aggregated features, 
which were then modulated with the original input through element-wise multiplication:
\begin{equation}
  \widetilde{h}_i=GELU(h_i)\odot z_i
\end{equation}
where $\odot$ denotes element-wise multiplication.
Collecting all patch features within the left and right hemispheres yields:
\begin{equation}
\begin{aligned}
  Z^L &= \left[\widetilde{h}_1^L,\widetilde{h}_2^L,\cdots,\widetilde{h}_N^L\right]\in\mathbb{R}^{N\times D},\\
  Z^R &= \left[\widetilde{h}_1^R,\widetilde{h}_2^R,\cdots,\widetilde{h}_N^R\right]\in\mathbb{R}^{N\times D}
\end{aligned}
\end{equation}

\begin{figure}
	\includegraphics[width=\columnwidth]{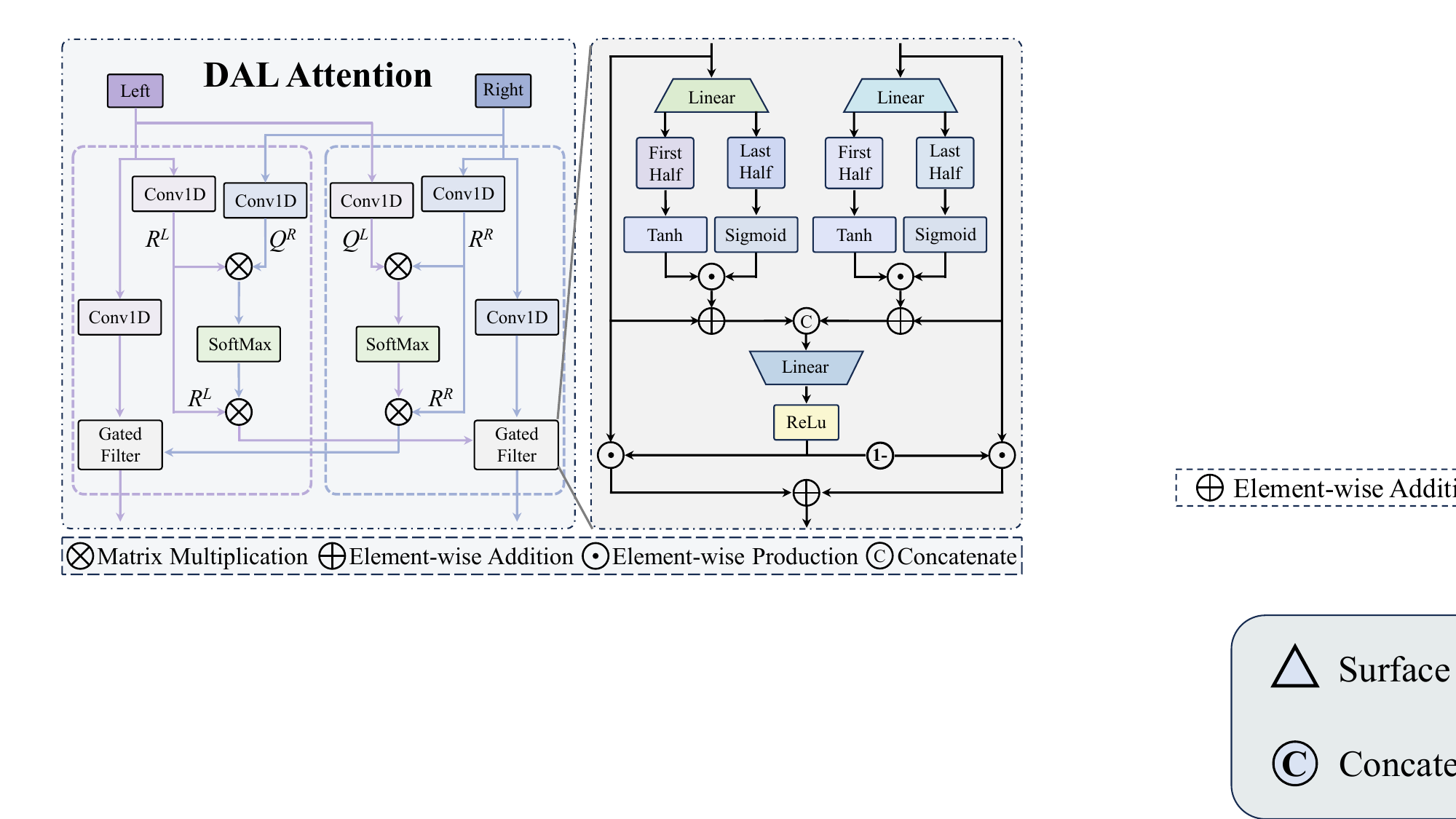}
	\caption{The detailed structures of the Dynamic Adaptive Lateralization-Aware Attention.}
    \label{fig6}
\end{figure}

\subsection{Dynamic Adaptive Lateralization-Aware Attention}
While structural lateralization differentiates the two hemispheres, the brain 
also maintains inter-hemispheric relationships through homotopic 
correspondences and complementary asymmetries. We introduced inter-hemispheric 
modeling to explicitly capture cross-hemispheric dependencies. To this end, 
we employed a cross-attention mechanism that selectively integrates information 
between hemispheres (\hyperref[fig6]{Fig. \ref*{fig6}}). By restricting attention 
to contralateral interactions, rather than all patches across the brain, we 
achieved computational efficiency while enabling the network to learn crucial 
asymmetry- and integration-related patterns.

Specifically, features of each hemisphere were projected into Embedding, Query, and Reference tokens:
\begin{align}
  \begin{aligned}
    E^L &= Conv_{1\times1}(Z^L), \\
    Q^L &= Conv_{1\times1}(Z^L), \\
    R^L &= Conv_{1\times1}(Z^L),
  \end{aligned}
  \qquad
  \begin{aligned}
    E^R &= Conv_{1\times1}(Z^R), \\
    Q^R &= Conv_{1\times1}(Z^R), \\
    R^R &= Conv_{1\times1}(Z^R)
  \end{aligned}
\end{align}
where the Reference token replaces both keys and values, reducing parameter 
redundancy and computational cost while enforcing a unified representational 
space for similarity measurement and information aggregation. This design is 
particularly beneficial for modeling cross-hemispheric relationships, where 
structural symmetry and homotopic correspondences are essential.

Cross-attention between hemispheres is formulated as:
\begin{equation}
\begin{aligned}
  \hat{Z}^L &= SoftMax\!\left(\frac{Q^L (R^R)^T}{\sqrt D}\right) R^R, \\
  \hat{Z}^R &= SoftMax\!\left(\frac{Q^R (R^L)^T}{\sqrt D}\right) R^L
\end{aligned}
\end{equation}
To further accommodate hemispheric lateralization of the brain, we designed a gated 
filter module that adaptively balanced hemispheric-specific and contralateral 
information. Query attends to its contralateral Reference feature were fused 
with original hemispheric features, allowing dominant intra-hemisphere 
characteristics to be preserved while adaptively incorporating complementary 
inter-hemispheric information. The process is formulated as:
\begin{equation}
\begin{aligned}
    \hat{Z}_1^L \in \mathbb{R}^{N \times D}, \;\hat{Z}_2^L \in \mathbb{R}^{N \times D} &= \mathrm{Split}(\mathrm{Linear}(\hat{Z}^L)) \\
    E_1^L \in \mathbb{R}^{N \times D}, \;E_2^L \in \mathbb{R}^{N \times D} &= \mathrm{Split}(\mathrm{Linear}(E^L))
\end{aligned}
\end{equation}
\vspace{-0.5em}  
\begin{equation}
\smash[b]{%
\begin{alignedat}{2}
W^L &= \mathrm{ReLU}\Big( 
  \mathrm{Linear}\big[ (\mathrm{Tanh}(\hat{Z}_1^L) \odot \mathrm{Sigmoid}(\hat{Z}_2^L) + \hat{Z}^L) \\
     &\quad \| (\mathrm{Tanh}(E_1^L) \odot \mathrm{Sigmoid}(E_2^L) + E^L) \big] 
\Big)
\end{alignedat}%
}
\end{equation}
\begin{gather}
\mathrm{Output}^L = W^L \odot \hat{Z}^L + (1 - W^L) \odot E^L
\end{gather}
By symmetry, the same process was applied to the other hemisphere to obtain 
$W^{R}$ and $Output^{R}$.

\subsection{Training Objectives}
Our framework is designed for fine-grained brain age prediction. 
The output layer consists of 161 parallel MLP heads: 
160 were assigned to predict the brain age of predefined region, 
and the remaining one predicted the global brain age. 
This design enables the model to capture both region-specific aging signatures and global aging trajectories.

Formally, for $n$ subjects and $m=160$ brain regions, let $\hat{y}_{i,j}$ and 
$y_{i,j}$ denote the global prediction and label. The region- and global-level losses are defined as:
\begin{gather}
  L_{region}=\frac{1}{n}\sum_{i=1}^{n}\frac{1}{m}\sum_{j=1}^{m}\left(\hat{y}_{i,j}-y_{i,j}\right)^2\\
  L_{global}=\frac{1}{n}\sum_{i=1}^n(\widehat{y}_i-y_i)^2
\end{gather}
To avoid fixed assigned or manually tuned weights, we adopt an 
uncertainty-based multi-task learning formulation\cite{kendall2018multi}. 
The final objective is expressed as:
\begin{equation}
  \begin{split}
  L_{total}=\frac{1}{2\sigma_{global}^2}L_{global}+\log\sigma_{global}+ \\ 
\frac{1}{2\sigma_{region}^2}L_{region}+\log\sigma_{region}
\end{split}
\end{equation}
where $\sigma_{global}$ and $\sigma_{region}$ are learnable parameters 
that representing task-dependent homoscedastic uncertainty of global and 
regional predictions, respectively.

This joint supervision encourages the model to integrate information across 
different scales: regional supervision promotes sensitivity to localized structural 
variations, while global supervision constrains the overall aging trajectory. 
By leveraging uncertainty-based weighting, the model adaptively balances the two 
objectives, yielding representations that reflect both fine-grained regional 
dynamics and the integrated process of brain maturation.

\renewcommand{\arraystretch}{0.8}
\begin{table*}[htbp]
\centering
\caption{Performance comparison of different networks.}
\label{tab2}
\resizebox{0.9\textwidth}{!}{
\begin{tabular}{cccccccc}
\toprule
\multirow{2}{*}{\textbf{\textit{Methods}}} & 
\multicolumn{1}{c}{\textbf{\textit{Global}}} & 
\multicolumn{1}{c}{\textbf{\textit{Global}}} & 
\multicolumn{1}{c}{\textbf{\textit{Global}}} & 
\multicolumn{1}{c}{\textbf{\textit{Region}}} & 
\multicolumn{1}{c}{\textbf{\textit{Region}}} & 
\multicolumn{1}{c}{\textbf{\textit{Region}}} & 
\multicolumn{1}{c}{\textbf{\textit{Params.}}} \\
& \textbf{\textit{MAE}} & \textbf{\textit{PCC}} & \textbf{\textit{SRCC}} & \textbf{\textit{MAE}} & \textbf{\textit{PCC}} & \textbf{\textit{SRCC}} & \textbf{(M)} \\ 
\toprule
\textbf{MoNet}              & 0.66±0.47 & 0.99±0.00 & 0.96±0.01 & - & - & - & -  \\
\textbf{S2CNN}              & 0.67±0.49 & 0.99±0.00 & 0.96±0.01 & - & - & - & -  \\
\textbf{ChebNet}            & 0.67±0.54 & 0.99±0.00 & 0.96±0.01 & - & - & - & -  \\
\textbf{GConvNet}           & 0.84±0.82 & 0.98±0.01 & 0.91±0.03 & - & - & - & -  \\
\textbf{PointNet++}         & 0.59±0.51 & 0.99±0.00 & 0.97±0.01 & - & - & - & -  \\
\textbf{Spherical UNet}     & 0.80±0.84 & 0.98±0.01 & 0.92±0.03 & - & - & - & -  \\
\toprule
\textbf{HRINet/1}           & 0.60±0.45 & 0.99±0.00 & 0.97±0.01 & - & - & - & 9.87 \\
\textbf{SiT-Small/1}        & 0.63±0.52 & 0.99±0.00 & 0.96±0.01 & - & - & - & 21.99 \\
\textbf{SiM-Small/1}        & 0.59±0.44 & 0.99±0.00 & 0.96±0.01 & - & - & - & 23.75 \\
\toprule
\textbf{HRINet/2}           & 0.59±0.47 & 0.99±0.00 & 0.97±0.01 & - & - & - & 9.45 \\
\textbf{SiT-Small/2}        & 0.60±0.51 & 0.99±0.00 & 0.96±0.01 & - & - & - & 21.70 \\
\textbf{SiM-Small/2}        & 0.57±0.48 & 0.99±0.00 & 0.97±0.01 & - & - & - & 23.46 \\
\toprule
\textbf{HRINet/3}           & \textit{OOM} & \textit{OOM} & \textit{OOM} & - & - & - & 9.72 \\
\textbf{SiT-Small/3}        & 0.55±0.48 & 0.99±0.00 & 0.97±0.01 & - & - & - & 22.32 \\
\textbf{SiM-Small/3}        & 0.53±0.43 & 0.99±0.00 & 0.97±0.01 & - & - & - & 24.08 \\
\toprule
\textbf{PVTv1-Tiny}         & 0.60±0.47 & 0.99±0.00 & 0.96±0.01 & 0.59±0.02 & 0.99±0.00 & 0.96±0.01 & 13.77 \\
\textbf{PVTv2-B1}           & 0.57±0.47 & 0.99±0.00 & 0.97±0.01 & 0.58±0.03 & 0.99±0.00 & 0.97±0.01 & 13.61 \\
\textbf{PoolFormer-S12}     & 0.68±0.58 & 0.99±0.00 & 0.95±0.01 & 0.80±0.05 & 0.98±0.01 & 0.94±0.01 & 12.53 \\
\textbf{MS-SiT}             & 0.51±0.38 & 0.99±0.00 & 0.97±0.01 & 0.56±0.04 & 0.99±0.00 & 0.97±0.01 & 27.64 \\
\textbf{SurfAge-Net}             & \textbf{0.45±0.39} & \textbf{0.99±0.00} & \textbf{0.98±0.01} & \textbf{0.54±0.04} & \textbf{0.99±0.00} & \textbf{0.97±0.01} & 14.08 \\
\bottomrule
\end{tabular}
}
\end{table*}

\subsection{Statistical Analysis}
Model performance was evaluated using the Pearson correlation coefficient (PCC) 
and Spearman rank correlation coefficient (SRCC) between predicted and 
chronological brain age, both globally and regionally. To obtain robust estimates 
of correlation strength and their variability, PCC and SRCC were calculated using 
a bootstrap procedure with 1,000 resamples, and the mean and standard deviation 
were reported. To assess regional deviations between predicted and chronological 
brain ages, paired t-tests were conducted for each of the 160 cortical region 
across subjects. To characterize region-specific cortical morphological 
development, two-sample t-tests were performed to compare morphological 
measures between extremely preterm or moderate-to-late preterm groups 
with the term-born neonates across the 160 cortical regions, with sex 
and postnatal age at scan included as covariates. Multiple comparisons 
were corrected using the Benjamini-Hochberg 
false discovery rate (FDR) procedure (\textit{q} $<$ 0.05).

\begin{table*}[ht]
\large
\centering
\caption{Quantitative analysis of module ablation experiments.}
\label{tab3}
\resizebox{\textwidth}{!}{
\begin{tabular}{cccccccccccc}
\toprule
\multirow{2}{*}{\textbf{\textit{Model}}} & 
\multirow{2}{*}{\textbf{\textit{SM}}} & 
\multirow{2}{*}{\textbf{\textit{CM}}} & 
\multirow{2}{*}{\textbf{\textit{CA}}} & 
\multirow{2}{*}{\textbf{\textit{GF}}} & 
\multicolumn{1}{c}{\textbf{\textit{Global}}} & 
\multicolumn{1}{c}{\textbf{\textit{Global}}} & 
\multicolumn{1}{c}{\textbf{\textit{Global}}} & 
\multicolumn{1}{c}{\textbf{\textit{Region}}} & 
\multicolumn{1}{c}{\textbf{\textit{Region}}} & 
\multicolumn{1}{c}{\textbf{\textit{Region}}} & 
\multicolumn{1}{c}{\textbf{\textit{Params.}}} \\
& & & & & \textbf{\textit{MAE}} & \textbf{\textit{PCC}} & \textbf{\textit{SRCC}} & \textbf{\textit{MAE}} & \textbf{\textit{PCC}} & \textbf{\textit{SRCC}} & \textbf{(M)} \\ 
\toprule
\textbf{(a)}    & \xmark & \xmark & \cmark  & \xmark & 0.69±0.56 & 0.99±0.00 & 0.96±0.01 & 0.71±0.07 & 0.99±0.00 & 0.96±0.01 & 6.75  \\
\textbf{(b)}    & \xmark & \cmark  & \cmark  & \xmark & 0.54±0.54 & 0.99±0.00 & 0.97±0.02 & 0.64±0.03 & 0.99±0.00 & 0.96±0.02 & 7.84  \\
\textbf{(c)}    & \cmark  & \xmark & \cmark  & \xmark & 0.54±0.48 & 0.99±0.00 & 0.97±0.01 & 0.62±0.01 & 0.99±0.00 & 0.97±0.01 & 11.42  \\
\textbf{(d)}    & \xmark & \xmark & \cmark  & \cmark  & 0.64±0.54 & 0.99±0.00 & 0.96±0.01 & 0.65±0.02 & 0.99±0.00 & 0.96±0.01 & 8.32  \\
\textbf{(e)}    & \xmark & \cmark  & \cmark  & \cmark  & 0.53±0.46 & 0.99±0.00 & 0.98±0.01 & 0.62±0.05 & 0.99±0.00 & 0.96±0.01 & 9.42  \\
\textbf{(f)}    & \cmark  & \xmark & \cmark  & \cmark  & 0.58±0.48 & 0.99±0.00 & 0.96±0.01 & 0.63±0.01 & 0.99±0.00 & 0.96±0.01 & 12.99  \\
\textbf{(g)}    & \xmark & \cmark  & \xmark & \xmark & 0.56±0.45 & 0.99±0.00 & 0.97±0.01 & 1.12±0.11 & 0.97±0.01 & 0.89±0.03 & 5.88 \\
\textbf{(h)}    & \cmark  & \xmark & \xmark & \xmark & 0.57±0.47 & 0.99±0.00 & 0.97±0.01 & 0.65±0.06 & 0.99±0.00 & 0.95±0.02 & 9.45 \\
\textbf{(i)}    & \cmark  & \cmark  & \xmark & \xmark & 0.54±0.43 & 0.99±0.00 & 0.97±0.01 & 0.61±0.01 & 0.99±0.00 & 0.96±0.01 & 10.55 \\
\textbf{(j)}    & \cmark  & \cmark  & \cmark  & \xmark & 0.52±0.46 & 0.99±0.00 & 0.96±0.01 & 0.60±0.01 & 0.99±0.00 & 0.96±0.01 & 12.51 \\
\textbf{(k)}    & \cmark  & \cmark  & \cmark  & \cmark  & \textbf{0.45±0.39} & \textbf{0.99±0.00} & \textbf{0.98±0.01} & \textbf{0.54±0.04} & \textbf{0.99±0.00} & \textbf{0.97±0.01} & 14.08 \\ 
\bottomrule
\end{tabular}
}
\end{table*}

\section{Results}

\subsection{Implementation Details}
Our method was implemented in Python 3.8.20 using the PyTorch library, 
and all experiments were conducted on a single NVIDIA A100 GPU. The input data were normalized 
to have zero mean and unit variance. All models were trained from scratch for 1200 
epochs using the AdamW optimizer $\beta_{1}=0.9$, $\beta_{2}=0.999$ with a weight decay of 1$\times$$10^{-8}$ and a batch 
size of 32. The initial learning rate was set to 5$\times$$10^{-5}$, scheduled 
with a 20-epoch linear warmup followed by cosine decay to a minimum of 5$\times$$10^{-6}$.

\subsection{Fine-Grained Infant Brain Age Prediction}
To comprehensively evaluate the effectiveness of the proposed SurfAge-Net, we conducted 
comparative experiments on \textit{dHCP}, benchmarking it against both GDL-based methods and 
recent attention- and Mamba-based architectures. The results are summarized in \hyperref[tab2]{Table \ref*{tab2}}.

We first examined global brain age prediction. Among the GDL models, 
PointNet++\cite{qi2017pointnet++} achieved the lowest MAE of 0.59, outperforming other GDL baselines 
whose MAEs exceeded 0.66. Attention-based methods such as 
HRINet\cite{zhao2024attention} and SiT\cite{dahan2022surface}, as well as the state-space-based 
SiM\cite{he2025surface} model, demonstrated superior performance over GDL models. Notably, SiM 
configured with an Ico-3 grid achieved the best performance among these methods, 
reaching an MAE of 0.53.

We then examined fine-grained prediction, in which models simultaneously estimated 
both global and region-level brain ages. SurfAge-Net achieved the best performance 
across all evaluation metrics, obtaining a whole-brain MAE of 0.45 and a region-level 
MAE of 0.54. Moreover, it achieved the highest PCC and SRCC, indicating stronger 
linear correlation and rank-order consistency with chronological age. Compared with 
MS-SiT\cite{dahan2024multiscale}—the closest competitor of SurfAge-Net that attained a whole-brain MAE of 0.51 
and a region-level MAE of 0.56—SurfAge-Net not only delivered higher prediction 
performance but also did so with approximately half the number of parameters. 
Other lightweight attention-based architectures, including 
PVTv1-Tiny\cite{wang2021pyramid}, PVTv2-B1\cite{wang2022pvt}, and PoolFormer-S12\cite{yu2022metaformer}, 
had comparable model sizes to SurfAge-Net but exhibited poorer predictive performance.

\begin{table*}[htbp]
\centering
\caption{Quantitative analysis of pooling methods ablation experiments.}
\label{tab4}
\resizebox{0.85\textwidth}{!}{
\begin{tabular}{cccccccccccc}
\toprule
\multicolumn{1}{c}{\textbf{\textit{Pooling}}} & 
\multicolumn{1}{c}{\textbf{\textit{Global}}} & 
\multicolumn{1}{c}{\textbf{\textit{Global}}} & 
\multicolumn{1}{c}{\textbf{\textit{Global}}} & 
\multicolumn{1}{c}{\textbf{\textit{Region}}} & 
\multicolumn{1}{c}{\textbf{\textit{Region}}} & 
\multicolumn{1}{c}{\textbf{\textit{Region}}} \\
\textbf{\textit{Strategy}} & \textbf{\textit{MAE}} & \textbf{\textit{PCC}} & \textbf{\textit{SRCC}} & \textbf{\textit{MAE}} & \textbf{\textit{PCC}} & \textbf{\textit{SRCC}} \\ 
\toprule
\textbf{\textit{w/o SM}}             & 0.53±0.46 & 0.99±0.00 & 0.98±0.01 & 0.62±0.05 & 0.99±0.00 & 0.96±0.01 \\
\textbf{\textit{Max pool}}           & 0.51±0.40 & 0.99±0.00 & 0.98±0.01 & 0.55±0.02 & 0.99±0.00 & 0.97±0.01 \\
\textbf{\textit{Min pool}}           & 0.53±0.45 & 0.99±0.00 & 0.98±0.01 & 0.55±0.01 & 0.99±0.00 & 0.97±0.01 \\
\textbf{\textit{Mean pool}}          & 0.52±0.41 & 0.99±0.00 & 0.98±0.01 & 0.58±0.05 & 0.99±0.00 & 0.97±0.01 \\
\textbf{\textit{Weighted pool}}      & 0.51±0.43 & 0.99±0.00 & 0.98±0.01 & 0.59±0.02 & 0.99±0.00 & 0.96±0.01 \\
\textbf{\textit{Stochastic pool}}    & \textbf{0.45±0.39} & \textbf{0.99±0.00} & \textbf{0.98±0.01} & \textbf{0.54±0.04} & \textbf{0.99±0.00} & \textbf{0.97±0.01} \\ 
\bottomrule
\end{tabular}
}

\end{table*}

\begin{table*}[htbp]
\centering

\captionsetup{justification=raggedright, singlelinecheck=false}

\caption{Quantitative analysis of subspace number ablation experiments.}
\label{tab5}
\resizebox{0.8\textwidth}{!}{
\begin{tabular}{cccccccccccc}
\toprule
\multirow{2}{*}{\textbf{\textit{k}}} & 
\multicolumn{1}{c}{\textbf{\textit{Global}}} & 
\multicolumn{1}{c}{\textbf{\textit{Global}}} & 
\multicolumn{1}{c}{\textbf{\textit{Global}}} & 
\multicolumn{1}{c}{\textbf{\textit{Region}}} & 
\multicolumn{1}{c}{\textbf{\textit{Region}}} & 
\multicolumn{1}{c}{\textbf{\textit{Region}}} & 
\multicolumn{1}{c}{\textbf{\textit{Params.}}} \\
& \textbf{\textit{MAE}} & \textbf{\textit{PCC}} & \textbf{\textit{SRCC}} & \textbf{\textit{MAE}} & \textbf{\textit{PCC}} & \textbf{\textit{SRCC}} & \textbf{(M)} \\ 
\toprule
\textbf{1}    & 0.55±0.46 & 0.99±0.00 & 0.96±0.01 & 0.65±0.02 & 0.99±0.00 & 0.96±0.01 & 14.09 \\
\textbf{2}    & 0.58±0.45 & 0.99±0.00 & 0.96±0.01 & 0.61±0.02 & 0.99±0.00 & 0.96±0.01 & 14.09 \\
\textbf{3}    & \textbf{0.45±0.39} & \textbf{0.99±0.00} & \textbf{0.98±0.01} & \textbf{0.54±0.04} & \textbf{0.99±0.00} & \textbf{0.97±0.01} & \textbf{14.08} \\ 
\textbf{4}    & 0.60±0.49 & 0.99±0.00 & 0.95±0.01 & 0.70±0.04 & 0.99±0.00 & 0.94±0.02 & 14.08 \\
\bottomrule
\end{tabular}
}

\end{table*}

\begin{table*}[ht]
\centering
\caption{Quantitative analysis of loss function ablation experiments.}
\label{tab6}
\resizebox{\textwidth}{!}{
\begin{tabular}{cccccccccccc}
\toprule
\multirow{2}{*}{\textbf{\textit{Loss item}}} & 
\multicolumn{1}{c}{\textbf{\textit{Global}}} & 
\multicolumn{1}{c}{\textbf{\textit{Global}}} & 
\multicolumn{1}{c}{\textbf{\textit{Global}}} & 
\multicolumn{1}{c}{\textbf{\textit{Region}}} & 
\multicolumn{1}{c}{\textbf{\textit{Region}}} & 
\multicolumn{1}{c}{\textbf{\textit{Region}}} & 
\multicolumn{1}{c}{\textbf{\textit{Params.}}} \\
& \textbf{\textit{MAE}} & \textbf{\textit{PCC}} & \textbf{\textit{SRCC}} & \textbf{\textit{MAE}} & \textbf{\textit{PCC}} & \textbf{\textit{SRCC}} & \textbf{(M)} \\ 
\toprule
\textbf{$L_{region}$}    & - & - & - & 0.78±0.05 & 0.98±0.01 & 0.93±0.02 & 14.08 \\
\textbf{$L_{global}$}    & 0.79±0.84 & 0.98±0.01 & 0.91±0.03 & - & - & - & 14.02 \\
\textbf{$L_{global}$+$L_{region}$}    & 0.52±0.41 & 0.99±0.00 & 0.97±0.01 & 0.60±0.02 & 0.99±0.00 & 0.97±0.01 & 14.08 \\
\textbf{$L_{total}$}    & \textbf{0.45±0.39} & \textbf{0.99±0.00} & \textbf{0.98±0.01} & \textbf{0.54±0.04} & \textbf{0.99±0.00} & \textbf{0.97±0.01} & \textbf{14.08} \\ 
\bottomrule
\end{tabular}
}
\end{table*}

\subsection{Ablation Experiments}

\subsubsection{Effects of Different Module}
To assess the contribution of each component, we performed ablation experiments 
by systematically combining the four core modules: the Spatial Mixer (SM), 
Channel Mixer (CM), Cross-Attention (CA), and Gated Filter (GF). Configurations 
were denoted as models (a)-(k), shown in \hyperref[tab3]{Table \ref*{tab3}}.

Compared with the baseline model (a), models (b) and (c) exhibit consistent 
improvements across all evaluation metrics, demonstrating that both SM and CM 
effectively enhance predictive performance. Likewise, model (d) outperformed 
model (a) on all metrics, confirming the effectiveness of the GF module. When SM 
or CM was further added into the CA+GF configuration (models (e) and (f)), 
additional performance gains were observed, indicating that these modules provided 
complementary benefits.

Interestingly, model (g), which include only the CM module, performed well on 
global measures but relatively poor on regional metrics. This pattern, corroborated 
by the superior regional performance of models (b) and (i), suggested that CM 
primarily enhances intra-regional feature representations while lacking the ability 
to capture inter-regional dependencies. Finally, from models (h) to (k), we 
observed a progressive improvement as more modules are incorporated, with the 
best performance achieved when all four modules are integrated. These results 
highlighted the synergistic interaction among SM, CM, CA and GF, confirming the 
necessity of each component in achieving optimal model performance.

\subsubsection{Effects of Different Pooling Methods}
Since the SM aggregates intra-hemispheric information by pooling patch 
features into a global representation, the choice of pooling strategy 
critically determines the quality of this representation. We compared 
several pooling methods, including max pooling, min pooling, mean pooling, 
weighted pooling and stochastic pooling, along with a variant where the 
SM was entirely removed.

According to \hyperref[tab4]{Table \ref*{tab4}}, the results showed that stochastic pooling achieved the best overall performance, 
whereas removing the SM led to the worst outcomes. This finding confirmed the 
importance of explicitly modeling intra-hemispheric interactions for accurate brain 
age prediction. Among the deterministic pooling methods, max pooling performed 
relatively better, suggesting that salient features played an important role in 
capturing hemispheric characteristics. In contrast, min pooling yielded the weakest 
result, likely due to its bias toward less informative features. Weighted and mean 
pooling exhibited moderate performance, but both remained inferior to stochastic 
pooling. These findings indicated that stochastic pooling is particularly 
well-suited for constructing robust hemispheric representations. From a biological 
perspective, this aligns with the fact that neuronal communication is inherently 
probabilistic rather than strictly deterministic\cite{otmakhov1993measuring,aitchison2016hamiltonian}. 
The stochastic mechanism may thus better capture the uncertainty and variability 
underlying intra-hemispheric information integration in the developing brain.

\subsubsection{Effects of Different Number of Subspace}
The CM enhanced patch-level representations by uniformly splitting each feature 
vector into $k$ subspaces, analogous to the multi-head design in attention mechanisms. 
This partitioning enables the model to learn complementary sub-representations from 
different feature subsets. To access the impact of this design, we evaluated 
different settings of $k\in\{1,2,3,4\}$, shown in \hyperref[tab5]{Table \ref*{tab5}}.

The results revealed that dividing features into three subspaces yielded 
the best overall performance, while four subspaces produced the worst 
outcomes. Configurations with one or two subspaces showed intermediate 
performance. These findings suggested a trade-off between representational 
diversity and per-subspace capacity. Specifically, increasing the number 
of subspaces enriches diversity by promoting disentangled learning, but 
simultaneously reduces the dimensionality of each subspace, potentially 
weakening its modeling capability. In our experiments, employing only 
one or two subspaces limited the ability of model to disentangle 
heterogeneous information, thereby constraining its expressive power. 
In contrast, using four subspaces might excessively fragmented the 
feature space, leading to unstable optimization and degraded performance. 
The three-subspace configuration appeared to achieve an optimal balance, 
offering sufficient diversity to capture complementary aging-related 
patterns while maintaining adequate feature dimensionality for effective 
representation learning.

\begin{table*}[htbp]
\centering
\caption{Generalization validation on images of term-born neonates in GSMCH.}
\label{tab7}
\resizebox{0.9\textwidth}{!}{
\begin{tabular}{cccccccccccc}
\toprule
\multirow{2}{*}{\textbf{\textit{Methods}}} & 
\multicolumn{1}{c}{\textbf{\textit{Global}}} & 
\multicolumn{1}{c}{\textbf{\textit{Global}}} & 
\multicolumn{1}{c}{\textbf{\textit{Global}}} & 
\multicolumn{1}{c}{\textbf{\textit{Region}}} & 
\multicolumn{1}{c}{\textbf{\textit{Region}}} & 
\multicolumn{1}{c}{\textbf{\textit{Region}}} \\
& \textbf{\textit{MAE}} & \textbf{\textit{PCC}} & \textbf{\textit{SRCC}} & \textbf{\textit{MAE}} & \textbf{\textit{PCC}} & \textbf{\textit{SRCC}} \\ 
\toprule
\textbf{PVTv1-Tiny}           & 1.18±0.70 & 0.67±0.17 & 0.59±0.28 & 1.19±0.11 & 0.66±0.17 & 0.69±0.21 \\
\textbf{PVTv2-B1}           & 1.59±0.80 & 0.73±0.17 & 0.82±0.17 & 1.51±0.17 & 0.71±0.17 & 0.79±0.16 \\
\textbf{PoolFormer-S12}          & 1.26±0.94 & 0.43±0.30 & 0.38±0.34 & 2.08±0.98 & 0.25±0.27 & 0.25±0.30 \\
\textbf{MS-SiT}        & 1.82±0.81 & 0.74±0.09 & 0.83±0.12 & 1.85±0.14 & 0.62±0.14 & 0.66±0.19 \\
\textbf{SurfAge-Net}    & \textbf{0.63±0.51} & \textbf{0.83±0.07} & \textbf{0.62±0.20} & \textbf{0.92±0.20} & \textbf{0.75±0.11} & \textbf{0.78±0.14} \\ 
\bottomrule
\end{tabular}
}
\end{table*}

\begin{table*}[htbp]
\centering
\caption{Generalization Validation on fetal images in LZUSH.}
\label{tab8}
\resizebox{0.9\textwidth}{!}{
\begin{tabular}{cccccccccccc}
\toprule
\multirow{2}{*}{\textbf{\textit{Methods}}} & 
\multicolumn{1}{c}{\textbf{\textit{Global}}} & 
\multicolumn{1}{c}{\textbf{\textit{Global}}} & 
\multicolumn{1}{c}{\textbf{\textit{Global}}} & 
\multicolumn{1}{c}{\textbf{\textit{Region}}} & 
\multicolumn{1}{c}{\textbf{\textit{Region}}} & 
\multicolumn{1}{c}{\textbf{\textit{Region}}} \\
& \textbf{\textit{MAE}} & \textbf{\textit{PCC}} & \textbf{\textit{SRCC}} & \textbf{\textit{MAE}} & \textbf{\textit{PCC}} & \textbf{\textit{SRCC}} \\ 
\toprule
\textbf{PVTv1-Tiny}         & 1.28±1.27 & 0.89±0.04 & 0.85±0.06 & 1.27±0.05 & 0.88±0.04 & 0.85±0.06 \\
\textbf{PVTv2-B1}           & 1.10±1.16 & 0.88±0.05 & 0.83±0.07 & 1.13±0.04 & 0.88±0.05 & 0.83±0.07 \\
\textbf{PoolFormer-S12}     & 1.77±1.01 & 0.96±0.01 & 0.96±0.02 & 2.58±0.59 & 0.76±0.07 & 0.76±0.08 \\
\textbf{MS-SiT}             & 1.26±1.15 & 0.90±0.04 & 0.86±0.06 & 1.36±0.09 & 0.88±0.04 & 0.84±0.06 \\
\textbf{SurfAge-Net}             & \textbf{1.02±0.66} & \textbf{0.96±0.01} & \textbf{0.96±0.01} & \textbf{1.10±0.11} & \textbf{0.94±0.01} & \textbf{0.94±0.03} \\ 
\bottomrule
\end{tabular}
}
\end{table*}

\subsubsection{Effects of Different Designs of Loss}
To evaluate the impact of supervision strategies, we compared four loss designs: 
(1) using only the mean regional MSE, (2) using only the global MSE, (3) directly 
summing the two losses, and (4) combining both losses with uncertainty-based learnable weights.

In \hyperref[tab6]{Table \ref*{tab6}}, the results showed that using 
either regional or global supervision alone led to suboptimal 
performance, with both single-loss variants yielding comparable 
results. Incorporating both losses (design 3) yielded a noticeable 
improvement, demonstrating that joint supervision enabled the model 
to capture complementary information from fine-grained regional 
patterns and overall brain-level developmental trajectories. The 
best performance was achieved by the uncertainty-weighted 
combination (design 4), consistently outperforming all other 
configurations across all evaluation metrics, which 
highlighted the importance of adaptive loss balancing, as the relative 
contributions of local and global information may vary throughout 
training. By allowing the model to dynamically adjust this balance, 
the optimization process became more flexible and better aligned with 
the intrinsic multi-scale structure of brain development.

\begin{figure*}
  \centering
	\includegraphics[width=0.95\textwidth]{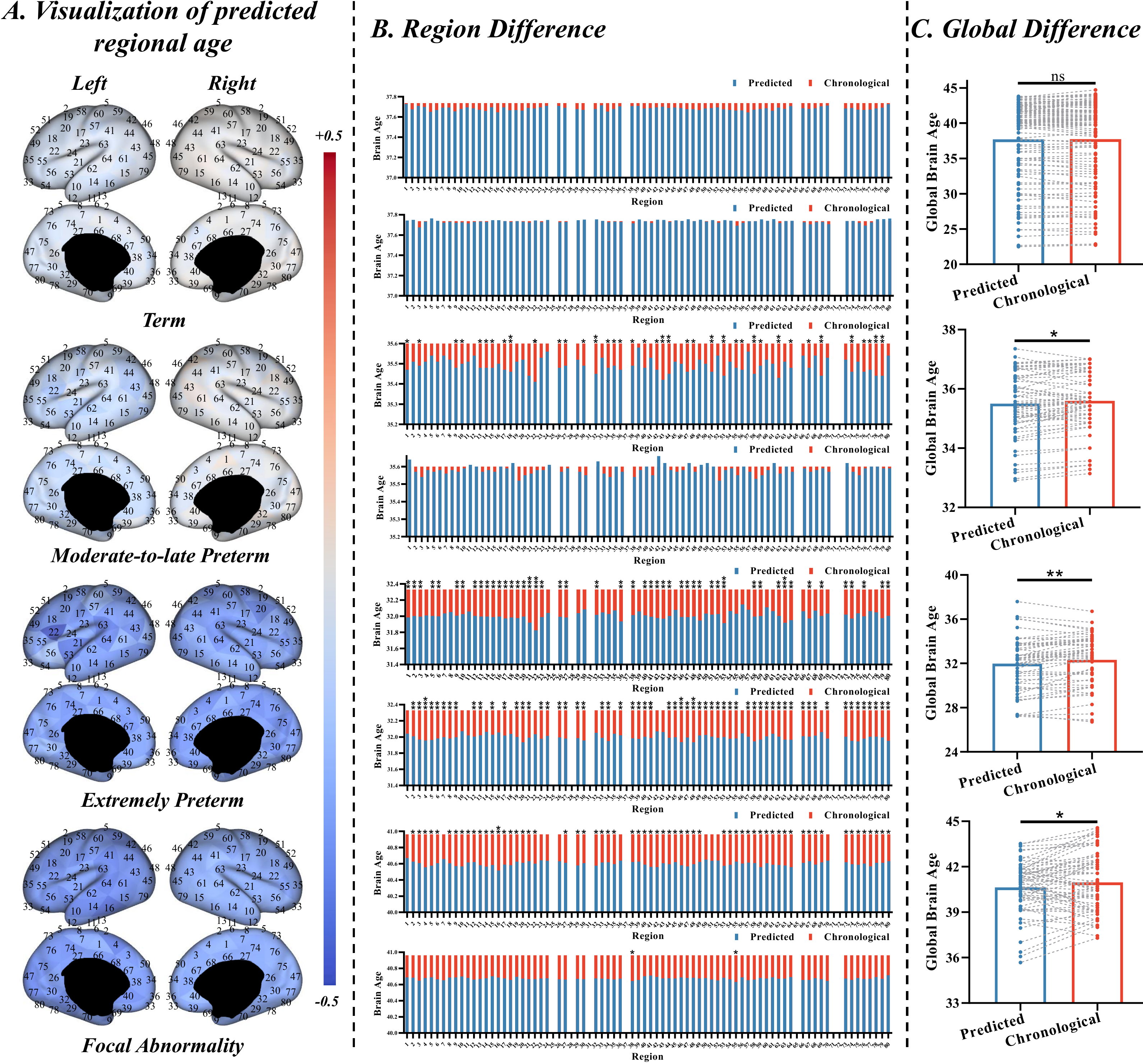}
	\caption{Regional and global brain developmental deviations in atypical developmental populations. (A) Visualization of predicted regional ages. Cortical surface maps intuitively illustrate the developmental status of each brain region, with regional indices annotated. Colors represent the deviation between predicted and chronological ages: positive values (e.g., +0.5) indicate accelerated development (0.5 weeks ahead), whereas negative values (e.g., -0.5) indicate delayed development (0.5 weeks behind). (B) Bar plots of regional differences. Comparisons between predicted and chronological ages across 80 cortical regions for the left and right hemispheres, respectively. Each bar corresponds to one region, enabling assessment of localized developmental deviations. *\textit{p} $<$ 0.05, **\textit{p} $<$ 0.01, ***\textit{p} $<$ 0.001 (paired t-test, FDR-corrected). (C) Global brain age differences. Paired-sample t-tests compare predicted global brain age with chronological age, highlighting overall developmental deviations in the atypical developmental cohort. Paired samples t-test: ns represents no signiﬁcance, *\textit{p} $<$ 0.05, **\textit{p} $<$ 0.01, ***\textit{p} $<$ 0.001 (paired-test).}
    \label{fig7}
\end{figure*}

\begin{figure*}
  \centering
	\includegraphics[width=0.9\textwidth]{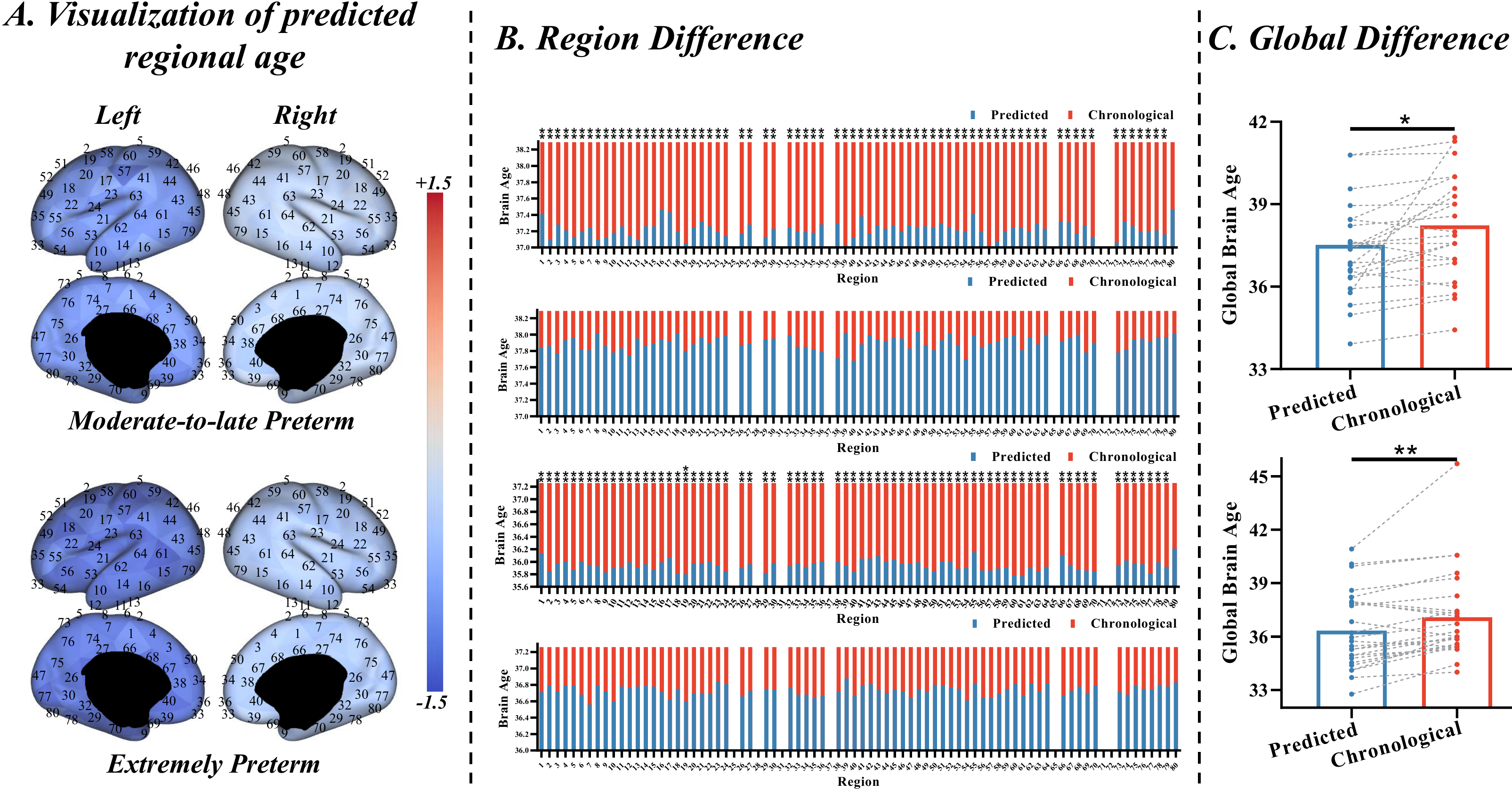}
	\caption{Regional and global brain developmental deviations in abnormal developmental populations from external datasets. (A) Visualization of predicted regional ages. Colors represent the deviation between predicted and chronological ages: positive values (e.g., +1.5) indicate accelerated development (1.5 weeks ahead), whereas negative values (e.g., -1.5) indicate delayed development (1.5 weeks behind). (B) Bar plots of regional differences. (C) Global brain age differences.}
    \label{fig8}
\end{figure*}

\subsection{Generalization Validation on Normative Cohort}
To assess the generalizability of different models, we evaluated their performance 
on images of term newborns in \textit{GSMCH} and fetuses in \textit{LZUSH}, respectively, as 
summarized in \hyperref[tab7]{Table \ref*{tab7}} and \hyperref[tab8]{Table \ref*{tab8}}. 
Although all models exhibited degradation 
across all metrics when transferred to the independent cohorts, SurfAge-Net 
consistently achieved the best performance, attaining the best generalization 
in both datasets. Notably, MS-SiT, which performed comparably to SurfAge-Net on \textit{dHCP}, 
showed a marked decline under distributional shifts, suggesting limited robustness 
to variations in population characteristics.

Interestingly, distinct patterns were observed between the two normative subsets. 
On term-born neonates in \textit{GSMCH}, lower MAE was accompanied by reduced PCC and SRCC. 
This likely reflected the very small sample size (16 infants), where random 
fluctuations can disproportionately affect correlation measures, resulting in small 
absolute errors but less stable rank ordering of predicted ages. Conversely, 
on \textit{LZUSH}, the MAE was higher, but both PCC and SRCC improved, likely due to the 
larger sample size (49 fetuses) could provide more reliable ordering and greater 
stability in correlations estimates. These results demonstrated that SurfAge-Net 
maintained superior generalizability across cohorts, even under substantial domain 
shifts, and that performance variations across subsets may reflect differences in 
sample size and label granularity rather than deficiencies in model robustness.

\subsection{Analysis on Atypical Developmental Infants}
\hyperref[fig7]{Fig \ref*{fig7}} illustrates both regional and global deviations in cortical maturation 
across term-born neonates and three atypical developmental cohorts. 
For the term-born neonates, cortical development appeared largely typical, 
with minimal regional deviations in cortical maturation (Panels A-B) and no 
regions exhibiting significant developmental delays or accelerations. The global 
brain age did not differ significantly from chronological age. 

For the moderate-to-late preterm group, cortical development was predominant 
delayed, particularly across the left hemisphere, which consistently lagged 
behind chronological age (Panels A-B). In contrast, several right-hemisphere 
regions, such as Superior Frontal Gyrus (region 1), Caudal Middle 
Frontal (region 18), and Parahippocampal (region 32), showed slight but 
nonsignificant overdevelopment. Marked regional developmental heterogeneity 
was evident (Panel B): while regions in the right hemisphere approximated 
normative age, several left-hemisphere regions exhibited delays of up to $\sim$0.2 
weeks. Except for the left hemisphere sensorimotor and superior parietal 
regions (regions 2, 4-8), the lateral temporal cortex (regions 11, 12), central 
sensorimotor strip (regions 41, 57, 60, 61, 63), insular 
cortex (regions 16, 19-21, 23, 24), and the frontoparietal 
cortex (regions 45, 46, 49, 50), most cortical regions showed significantly 
lower predicted brain age than chronological age, indicating widespread cortical 
maturation delay. The global brain age was also significantly lower than chronological age (Panel C).

For the extremely preterm group, more pronounced cortical delays were observed 
across both hemispheres (panel A), with an average delay of approximately 0.4 
weeks (panel B). Compared with the moderate-to-late preterm group, a larger 
number of regions exhibited significant developmental delays, accompanied by a 
more evident reduction in global brain age (\textit{p} $<$ 0.01, panel C). 

Similar to the two preterm cohorts, neonates with focal brain abnormality showed 
reduced brain age both globally and regionally (panel A-C), with an overall 
severity intermediate between the moderate-to-late and extremely preterm 
groups (panel B). In the left hemisphere, nearly all regions, except the 
sensorimotor regions (regions 23, 24, 26) and the frontal cortex (regions 50-52), 
showed significant developmental delays. In the right hemisphere, only the 
Superior Frontal Gyrus (region 38) and Pars Triangularis (region 55) exhibited 
comparable alterations. These findings highlighted the clinical relevance of 
fine-grained brain age prediction. Infants born at earlier gestational ages or 
presenting with focal brain abnormalities exhibited measurable developmental 
delays, suggesting the potential utility of SurfAge-Net-derived brain age as a 
promising biomarker for early risk stratification and personalized intervention 
planning to mitigate long-term neurodevelopmental deficits.

\subsection{External Validation on Atypical Developmental Infants}
To further assess the external validity of SurfAge-Net, we examined two atypical 
developmental populations from the \textit{GSMCH} datasets. Both groups showed significantly 
decreased predicted global and regional ages relative to chronological age (\hyperref[fig8]{Fig. \ref*{fig8}}), 
consistent with the results observed in the \textit{dHCP} datasets. For the moderate-to-late 
preterm group, widespread cortical developmental delays were found, with the left 
hemisphere exhibiting substantially greater delays than the right (approximately 0.7 
weeks on average, panel A-B). The average delay was about $\sim$1 week in the left 
hemisphere and $\sim$0.3 weeks in the right (panel B). Marked regional heterogeneity 
was also observed, as almost all regions in the left hemisphere exhibited pronounced 
maturation delays. The global brain age was significantly lower than chronological 
age (\textit{p} $<$ 0.05). For the extremely preterm group in \textit{GSMCH}, similar spatial and global 
patterns were observed, but with more pronounced severity (panels A-C). The left 
hemisphere exhibited comparable levels of delay, with the Precentral Gyrus (region 19) 
showing the most pronounced deficit. 

\begin{figure*}
  \centering
	\includegraphics[width=0.8\textwidth]{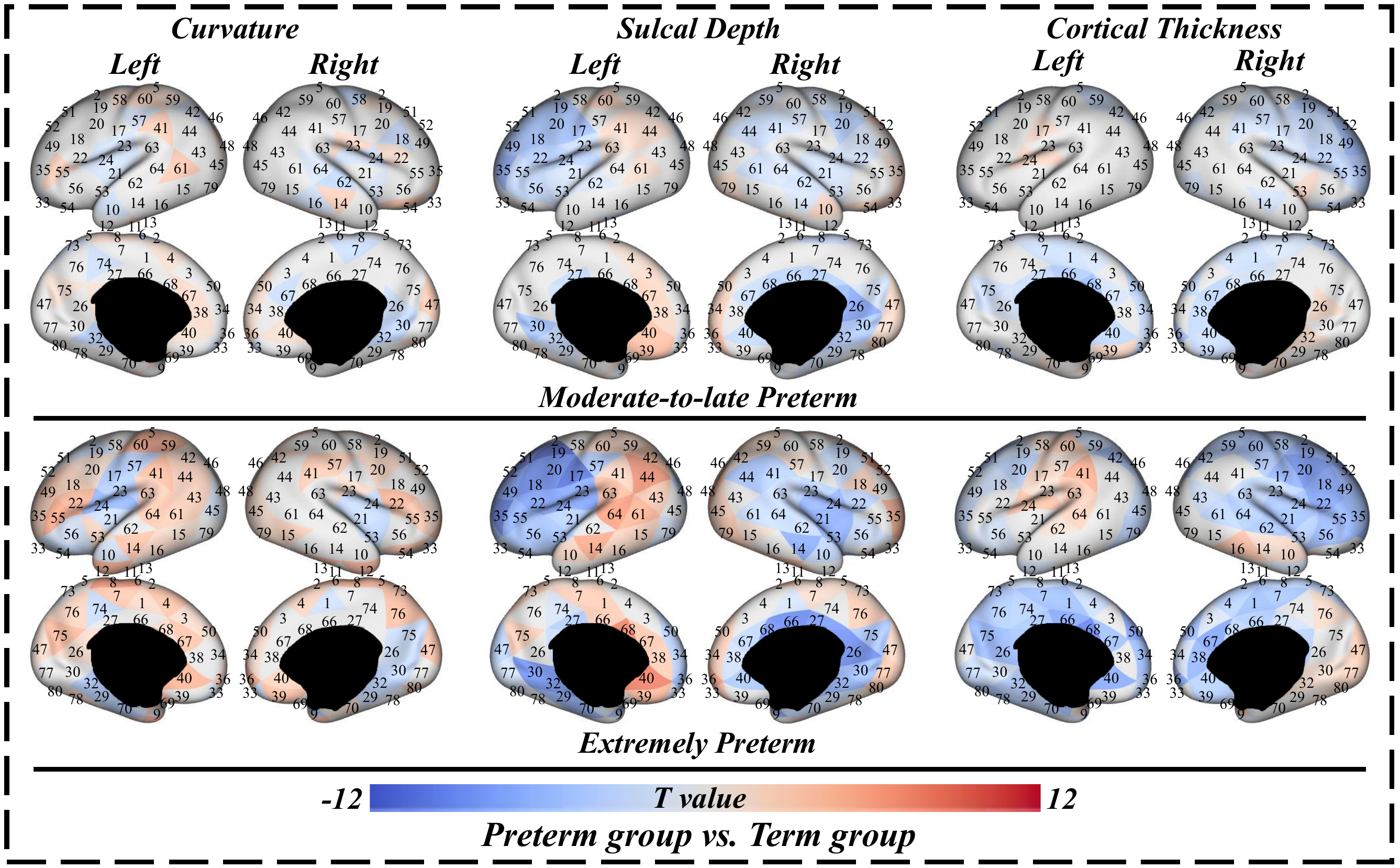}
	\caption{Regional differences in cortical morphology between preterm and term neonates. Multiple comparisons were corrected by the FDR method, and brain regions with \textit{q} $<$ 0.05 were visualized.}
    \label{fig9}
\end{figure*}

\subsection{Regional Deviation of Local Morphological Indicators}
To characterize region-specific cortical morphological development 
during the perinatal period, we examined how mean curvature, sulcal 
depth, and thickness vary across different cortical regions in preterm 
infants relative to term controls, and further investigated their 
associations with regional brain age gap. As observed in \hyperref[fig9]{Fig. \ref*{fig9}}, 
analysis of regional morphological deviations revealed distinct and 
feature-specific alteration patterns in preterm infants relative to 
term controls. Notably, sulcal depth and cortical thickness in the 
prefrontal and limbic cortices were significantly reduced in preterm 
infants, with the magnitude of reduction increasing with decreasing 
gestational age, indicating a dose-dependent effect of prematurity. 
In contrast, curvature exhibited a divergent pattern: prefrontal and 
temporo-parietal junction regions showed significantly increased 
curvature in extremely preterm infants, accompanied by elevated 
sulcal depth in parietal and medial cortical regions. Importantly, 
these curvature and sulcal depth increases were largely absent in 
the moderate-to-late preterm group, suggesting that such alterations 
are specific to more severe prematurity and are not captured by 
moderate deviations in gestational age. Notably, although the 
brain-age gap increased globally from moderate-to-late to extremely 
preterm infants, local cortical morphological changes did not converge 
toward a uniform direction. This inconsistency likely reflects the 
non-linear and region-specific nature of perinatal cortical development, 
during which measures such as curvature and thickness may follow 
inverted U-shaped or otherwise non-monotonic trajectories around 
the term-equivalent window\cite{xu2022spatiotemporal}. As a result, elevated or attenuated 
local morphological measures in extremely preterm infants may indicate 
altered developmental timing rather than a simple linear relationship 
with global brain-age delay.

\section{Discussion}
In this study, we introduced a novel patch-level brain age prediction framework, 
termed SurfAge-Net, which enables precise and region-specific assessment 
of cortical maturation in infancy. SurfAge-Net achieved substantial 
performance gains over existing methods in both global and regional 
brain-age prediction and demonstrated strong generalizability across 
multiple independent cohorts. By capturing fine-grained variations in 
cortical development, SurfAge-Net revealed complex and heterogeneous 
developmental patterns throughout the cortex that are often masked by 
coarser-scale approaches. Importantly, the model successfully identified 
subtle developmental delays in atypical neonate populations, including preterm and 
focal abnormality groups, highlighting its potential as a sensitive biomarker for 
early neurodevelopmental risks stratification. 

Unlike conventional whole-brain models that yield only a single global brain age, 
SurfAge-Net could delineate heterogeneous maturation degree within an individual brain, 
providing new insight into the dynamic and regionally asynchronous nature of early 
brain development. The fine-grained, patch-level framework further enables precise 
localization of developmental abnormalities at the cortical region level, 
identifying specific cortical regions that deviate from normative trajectories. 
This spatial resolution offers critical clinical value, revealing which cortical 
areas are most susceptible to atypical development—such as the sensorimotor, 
temporal, and frontoparietal cortices that are commonly affected in preterm 
infants\cite{kelly2024cortical,bouyssi2018altered}. In this study, this spatially 
resolved modeling revealed marked heterogeneity in cortical maturation in preterm 
infants, with accelerated development in prefrontal cortex and postcentral gyrus 
and delayed maturation in most of the high-order cortices, consistent with the 
hierarchical pattern of cortical development observed in previous 
studies\cite{dimitrova2021preterm,kelly2024cortical,ball2013development}, in which primary 
sensorimotor and visual areas mature earlier than high-order association areas. 
Such localization enhances both clinical interpretability and practicability of  
brain age modeling, facilitating early diagnosis and individualized intervention 
strategies tailored to specific cortical vulnerabilities. 

The human brain is organized as a complex, interconnected network rather than 
isolated regions. Morphometric similarity network studies have shown that brain 
regions with similar developmental trajectories tend to form tightly coupled networks, 
reflecting coordinated maturation across the 
cortex\cite{wang2024profiling,zheng2021developmental,seidlitz2018morphometric}. 
Therefore, predicting the developmental age of a single cortical patch necessitates 
integrating information not only from the target patch but also from anatomically 
and functionally related regions. In addition, brain asymmetry is a fundamental 
characteristic of early brain development, manifested as structural lateralization, 
differential growth rates, and asymmetric gene expression patterns between bilateral 
hemispheres\cite{liu2021diffusion,duboc2015asymmetry,kong2018mapping}. 
SurfAge-Net embodies these neurodevelopmental 
principles by explicitly modeling both intra-hemispheric dependencies and 
inter-hemispheric interactions, while accounting for hemispheric asymmetry 
in maturation. Within each hemisphere, the Spatial Mixer aggregates contextual 
information from related patches to capture coordinated morphological patterns, 
and the Channel Mixer adaptively reweights feature channels according to their 
contribution to brain age prediction, thereby enhancing informative signals 
while suppressing redundant or noisy signals. 
To model contralateral influences, the Dynamic Lateralization-Aware 
Attention mechanism selectively integrates 
information from the opposite hemisphere, regulated by gating filters 
that filter out irrelevant inputs. The improved performance of 
SurfAge-Net demonstrates that modeling these complex spatial 
relationships (i.e., ipsilateral heterogeneity and contralateral connectivity) is 
essential for accurately characterizing fine-grained, region-specific developmental 
trajectories. This also enables the model to reveal subtle, spatially organized 
maturational dynamics that traditional whole-brain methods often overlook. 
Moreover, SurfAge-Net consistently predicts younger brain ages in the left hemisphere 
compared to the right across multiple datasets, which is in line with established 
lateralization patterns in early brain development\cite{nam2015alterations}. 
This asymmetry likely corresponds to differential maturation rates and emerging 
functional specializations between hemispheres\cite{mento2010functional,schmitz2023altered}.

Our results provide strong evidence supporting the clinical interpretability of 
SurfAge-Net by demonstrating findings closely aligned with previous neuroimaging studies 
linking prematurity and focal lesions to disrupted cortical growth and delayed structural 
maturation\cite{dimitrova2021preterm,zhang2025dynamic}. 
The predicted brain age exhibits graded decline 
from term-born neonates to moderate-to-late preterm, and further to extremely preterm 
infants, reflecting a continuum of neurodevelopmental impairment that mirrors 
clinical observations of increasing gestational risk 
severity\cite{dimitrova2020heterogeneity,nivins2025gestational}. 
This progressive decrease in predicted brain age not only 
validates clinical reliability of SurfAge-Net, but also demonstrates its sensitivity 
to varying degrees of atypical developmental conditions. Furthermore, the spatial 
patterns of cortical maturation delays identified by SurfAge-Net correspond with 
established findings of regionally heterogeneous brain development in early 
infancy\cite{dimitrova2021preterm,liu2021diffusion}. Specifically, 
the model reveals pronounced delays in regions such as the prefrontal, 
lateral temporal, and sensorimotor cortices in extremely premature infants, 
consistent with diffuse cortical immaturity and region-specific vulnerability reported in the previous 
literature\cite{kelly2024cortical,kline2020early,hinojosa2017clinical}. 
The observed gradient—from 
relatively preserved maturation in moderate-to-late preterm infants to more 
widespread delays in extremely preterm cases—further confirms SurfAge-Net's ability to 
detect subtle regional variations during cortical development. These results 
indicate that SurfAge-Net not only reflects global developmental trends but also 
precisely maps region-specific alterations, thereby providing a biologically 
plausible and clinically interpretable framework for characterizing atypical cortical 
maturation in the developing brain.

Despite the promising results, there are several limitations that should be 
acknowledged. First, the sample size across the three datasets remain relatively 
limited, necessitating future validation on larger and diverse cohorts to ensure 
the generalizability and robustness of the proposed paradigm. Second, although our 
patch-level brain age prediction provides finer resolution than whole-brain 
prediction approaches, it may still be relatively coarse. Future studies could 
explore vertex-level or even higher-resolution predictions to achieve more precise 
characterization of cortical development. Third, while our focus was primarily on 
cortical features, other clinical variables such as genetic factors, perinatal 
complications, and environmental influences, were not incorporated. Integrating 
these multimodal data could further enhance prediction accuracy and clinical 
interpretability. Lastly, the present work focuses exclusively on early brain 
development; expanding this approach to other conditions like childhood development 
and aging could enable a more comprehensive evaluation of brain health across the lifespan.

\section{Conclusion}
We introduce SurfAge-Net, the first fine-grained, patch-level brain age 
prediction paradigm that achieves state-of-the-art performance across three 
datasets, with a global MAE of 0.54 and a regional MAE of 0.45.  SurfAge-Net 
facilitates precise and clinically interpretable mapping of atypical cortical 
development in clinical populations, including moderate-to-late preterm, extremely 
preterm, and focal abnormality infants. By providing a sensitive, region-specific 
insights into neurodevelopmental deviations, SurfAge-Net offers a powerful tool for 
early detection, risk stratification, and personalized intervention in pediatric 
neurological care.

\bibliographystyle{IEEEtran}
\bibliography{IEEEabrv,reference}

\begin{thebibliography}{10}
\providecommand{\url}[1]{#1}
\csname url@samestyle\endcsname
\providecommand{\newblock}{\relax}
\providecommand{\bibinfo}[2]{#2}
\providecommand{\BIBentrySTDinterwordspacing}{\spaceskip=0pt\relax}
\providecommand{\BIBentryALTinterwordstretchfactor}{4}
\providecommand{\BIBentryALTinterwordspacing}{\spaceskip=\fontdimen2\font plus
\BIBentryALTinterwordstretchfactor\fontdimen3\font minus \fontdimen4\font\relax}
\providecommand{\BIBforeignlanguage}[2]{{%
\expandafter\ifx\csname l@#1\endcsname\relax
\typeout{** WARNING: IEEEtran.bst: No hyphenation pattern has been}%
\typeout{** loaded for the language `#1'. Using the pattern for}%
\typeout{** the default language instead.}%
\else
\language=\csname l@#1\endcsname
\fi
#2}}
\providecommand{\BIBdecl}{\relax}
\BIBdecl

\bibitem{dehaene2015infancy}
G.~Dehaene-Lambertz and E.~S. Spelke, ``The infancy of the human brain,'' \emph{Neuron}, vol.~88, no.~1, pp. 93--109, 2015.

\bibitem{wilson2021development}
S.~Wilson, M.~Pietsch, L.~Cordero-Grande, A.~N. Price, J.~Hutter, J.~Xiao, L.~McCabe, M.~A. Rutherford, E.~J. Hughes, S.~J. Counsell \emph{et~al.}, ``Development of human white matter pathways in utero over the second and third trimester,'' \emph{Proceedings of the National Academy of Sciences}, vol. 118, no.~20, p. e2023598118, 2021.

\bibitem{franke2012brain}
K.~Franke, E.~Luders, A.~May, M.~Wilke, and C.~Gaser, ``Brain maturation: predicting individual brainage in children and adolescents using structural mri,'' \emph{Neuroimage}, vol.~63, no.~3, pp. 1305--1312, 2012.

\bibitem{whitmore2023brainage}
L.~B. Whitmore, S.~J. Weston, and K.~L. Mills, ``Brainage as a measure of maturation during early adolescence,'' \emph{Imaging Neuroscience}, vol.~1, pp. 1--21, 2023.

\bibitem{licht2009brain}
D.~J. Licht, D.~M. Shera, R.~R. Clancy, G.~Wernovsky, L.~M. Montenegro, S.~C. Nicolson, R.~A. Zimmerman, T.~L. Spray, J.~W. Gaynor, and A.~Vossough, ``Brain maturation is delayed in infants with complex congenital heart defects,'' \emph{The Journal of thoracic and cardiovascular surgery}, vol. 137, no.~3, pp. 529--537, 2009.

\bibitem{hazlett2017early}
H.~C. Hazlett, H.~Gu, B.~C. Munsell, S.~H. Kim, M.~Styner, J.~J. Wolff, J.~T. Elison, M.~R. Swanson, H.~Zhu, K.~N. Botteron \emph{et~al.}, ``Early brain development in infants at high risk for autism spectrum disorder,'' \emph{Nature}, vol. 542, no. 7641, pp. 348--351, 2017.

\bibitem{he2021global}
S.~He, P.~E. Grant, and Y.~Ou, ``Global-local transformer for brain age estimation,'' \emph{IEEE transactions on medical imaging}, vol.~41, no.~1, pp. 213--224, 2021.

\bibitem{zheng2023preterm}
W.~Zheng, X.~Wang, T.~Liu, B.~Hu, and D.~Wu, ``Preterm-birth alters the development of nodal clustering and neural connection pattern in brain structural network at term-equivalent age,'' \emph{Human Brain Mapping}, vol.~44, no.~16, pp. 5372--5386, 2023.

\bibitem{dimitrova2021preterm}
R.~Dimitrova, M.~Pietsch, J.~Ciarrusta, S.~P. Fitzgibbon, L.~Z. Williams, D.~Christiaens, L.~Cordero-Grande, D.~Batalle, A.~Makropoulos, A.~Schuh \emph{et~al.}, ``Preterm birth alters the development of cortical microstructure and morphology at term-equivalent age,'' \emph{NeuroImage}, vol. 243, p. 118488, 2021.

\bibitem{wang2024profiling}
Y.~Wang, D.~Zhu, L.~Zhao, X.~Wang, Z.~Zhang, B.~Hu, D.~Wu, and W.~Zheng, ``Profiling cortical morphometric similarity in perinatal brains: insights from development, sex difference, and inter-individual variation,'' \emph{NeuroImage}, vol. 295, p. 120660, 2024.

\bibitem{shaw2008neurodevelopmental}
P.~Shaw, N.~J. Kabani, J.~P. Lerch, K.~Eckstrand, R.~Lenroot, N.~Gogtay, D.~Greenstein, L.~Clasen, A.~Evans, J.~L. Rapoport \emph{et~al.}, ``Neurodevelopmental trajectories of the human cerebral cortex,'' \emph{Journal of neuroscience}, vol.~28, no.~14, pp. 3586--3594, 2008.

\bibitem{huang2022mapping}
Y.~Huang, Z.~Wu, F.~Wang, D.~Hu, T.~Li, L.~Guo, L.~Wang, W.~Lin, and G.~Li, ``Mapping developmental regionalization and patterns of cortical surface area from 29 post-menstrual weeks to 2 years of age,'' \emph{Proceedings of the National Academy of Sciences}, vol. 119, no.~33, p. e2121748119, 2022.

\bibitem{zheng2023spatiotemporal}
W.~Zheng, L.~Zhao, Z.~Zhao, T.~Liu, B.~Hu, and D.~Wu, ``Spatiotemporal developmental gradient of thalamic morphology, microstructure, and connectivity fromthe third trimester to early infancy,'' \emph{Journal of Neuroscience}, vol.~43, no.~4, pp. 559--570, 2023.

\bibitem{thomason2020development}
M.~E. Thomason, ``Development of brain networks in utero: relevance for common neural disorders,'' \emph{Biological psychiatry}, vol.~88, no.~1, pp. 40--50, 2020.

\bibitem{wu2024comparative}
X.~Wu, C.~Xie, F.~Cheng, Z.~Li, R.~Li, D.~Xu, H.~Kim, J.~Zhang, H.~Liu, and M.~Liu, ``Comparative evaluation of interpretation methods in surface-based age prediction for neonates,'' \emph{NeuroImage}, vol. 300, p. 120861, 2024.

\bibitem{dahan2022surface}
S.~Dahan, A.~Fawaz, L.~Z. Williams, C.~Yang, T.~S. Coalson, M.~F. Glasser, A.~D. Edwards, D.~Rueckert, and E.~C. Robinson, ``Surface vision transformers: Attention-based modelling applied to cortical analysis,'' in \emph{International Conference on Medical Imaging with Deep Learning}.\hskip 1em plus 0.5em minus 0.4em\relax PMLR, 2022, pp. 282--303.

\bibitem{li2025surfgnn}
Z.~Li, J.~Zhang, Y.~Zeng, J.~Lin, D.~Zhang, J.~Zhang, D.~Xu, H.~Kim, B.~Liu, and M.~Liu, ``Surfgnn: A robust surface-based prediction model with interpretability for coactivation maps of spatial and cortical features,'' \emph{Medical Image Analysis}, p. 103793, 2025.

\bibitem{cole2017predicting}
J.~H. Cole, R.~P. Poudel, D.~Tsagkrasoulis, M.~W. Caan, C.~Steves, T.~D. Spector, and G.~Montana, ``Predicting brain age with deep learning from raw imaging data results in a reliable and heritable biomarker,'' \emph{NeuroImage}, vol. 163, pp. 115--124, 2017.

\bibitem{dartora2024deep}
C.~Dartora, A.~Marseglia, G.~M{\aa}rtensson, G.~Rukh, J.~Dang, J.-S. Muehlboeck, L.-O. Wahlund, R.~Moreno, J.~Barroso, D.~Ferreira \emph{et~al.}, ``A deep learning model for brain age prediction using minimally preprocessed t1w images as input,'' \emph{Frontiers in Aging Neuroscience}, vol.~15, p. 1303036, 2024.

\bibitem{spalletta2018brain}
G.~Spalletta, F.~Piras, T.~Gili \emph{et~al.}, \emph{Brain morphometry}.\hskip 1em plus 0.5em minus 0.4em\relax Springer, 2018.

\bibitem{wang2023age}
Y.~Wang, Y.~Zhang, W.~Zheng, X.~Liu, Z.~Zhao, S.~Li, N.~Chen, L.~Yang, L.~Fang, Z.~Yao \emph{et~al.}, ``Age-related differences of cortical topology across the adult lifespan: Evidence from a multisite mri study with 1427 individuals,'' \emph{Journal of Magnetic Resonance Imaging}, vol.~57, no.~2, pp. 434--443, 2023.

\bibitem{bethlehem2022brain}
R.~A. Bethlehem, J.~Seidlitz, S.~R. White, J.~W. Vogel, K.~M. Anderson, C.~Adamson, S.~Adler, G.~S. Alexopoulos, E.~Anagnostou, A.~Areces-Gonzalez \emph{et~al.}, ``Brain charts for the human lifespan,'' \emph{Nature}, vol. 604, no. 7906, pp. 525--533, 2022.

\bibitem{xu2022spatiotemporal}
X.~Xu, C.~Sun, J.~Sun, W.~Shi, Y.~Shen, R.~Zhao, W.~Luo, M.~Li, G.~Wang, and D.~Wu, ``Spatiotemporal atlas of the fetal brain depicts cortical developmental gradient,'' \emph{Journal of Neuroscience}, vol.~42, no.~50, pp. 9435--9449, 2022.

\bibitem{sheng2024no}
Y.~Sheng, Y.~Wang, X.~Wang, Z.~Zhang, D.~Zhu, and W.~Zheng, ``No sex difference in maturation of brain morphology during the perinatal period,'' \emph{Brain Structure and Function}, vol. 229, no.~8, pp. 1979--1994, 2024.

\bibitem{he2025surface}
R.~He, W.~Zheng, L.~Zhao, Y.~Wang, D.~Zhu, D.~Wu, and B.~Hu, ``Surface vision mamba: Leveraging bidirectional state space model for efficient spherical manifold representation,'' in \emph{International Conference on Medical Image Computing and Computer-Assisted Intervention}.\hskip 1em plus 0.5em minus 0.4em\relax Springer, 2025, pp. 599--608.

\bibitem{zhao2024attention}
L.~Zhao, D.~Zhu, X.~Wang, X.~Liu, T.~Li, B.~Wang, Z.~Yao, W.~Zheng, and B.~Hu, ``An attention-based hemispheric relation inference network for perinatal brain age prediction,'' \emph{IEEE Journal of Biomedical and Health Informatics}, vol.~28, no.~8, pp. 4483--4493, 2024.

\bibitem{zheng2019multi}
W.~Zheng, T.~Eilam-Stock, T.~Wu, A.~Spagna, C.~Chen, B.~Hu, and J.~Fan, ``Multi-feature based network revealing the structural abnormalities in autism spectrum disorder,'' \emph{IEEE Transactions on Affective Computing}, vol.~12, no.~3, pp. 732--742, 2019.

\bibitem{sporns2004organization}
O.~Sporns, D.~R. Chialvo, M.~Kaiser, and C.~C. Hilgetag, ``Organization, development and function of complex brain networks,'' \emph{Trends in cognitive sciences}, vol.~8, no.~9, pp. 418--425, 2004.

\bibitem{chung2018use}
Y.~Chung, J.~Addington, C.~E. Bearden, K.~Cadenhead, B.~Cornblatt, D.~H. Mathalon, T.~McGlashan, D.~Perkins, L.~J. Seidman, M.~Tsuang \emph{et~al.}, ``Use of machine learning to determine deviance in neuroanatomical maturity associated with future psychosis in youths at clinically high risk,'' \emph{JAMA psychiatry}, vol.~75, no.~9, pp. 960--968, 2018.

\bibitem{becker2018gaussian}
B.~G. Becker, T.~Klein, C.~Wachinger, A.~D.~N. Initiative \emph{et~al.}, ``Gaussian process uncertainty in age estimation as a measure of brain abnormality,'' \emph{NeuroImage}, vol. 175, pp. 246--258, 2018.

\bibitem{hu2019hierarchical}
D.~Hu, Z.~Wu, W.~Lin, G.~Li, and D.~Shen, ``Hierarchical rough-to-fine model for infant age prediction based on cortical features,'' \emph{IEEE journal of biomedical and health informatics}, vol.~24, no.~1, pp. 214--225, 2019.

\bibitem{beheshti2022accuracy}
I.~Beheshti, N.~Maikusa, and H.~Matsuda, ``The accuracy of t1-weighted voxel-wise and region-wise metrics for brain age estimation,'' \emph{Computer Methods and Programs in Biomedicine}, vol. 214, p. 106585, 2022.

\bibitem{liu2022brain}
X.~Liu, I.~Beheshti, W.~Zheng, Y.~Li, S.~Li, Z.~Zhao, Z.~Yao, and B.~Hu, ``Brain age estimation using multi-feature-based networks,'' \emph{Computers in Biology and Medicine}, vol. 143, p. 105285, 2022.

\bibitem{huang2017age}
T.-W. Huang, H.-T. Chen, R.~Fujimoto, K.~Ito, K.~Wu, K.~Sato, Y.~Taki, H.~Fukuda, and T.~Aoki, ``Age estimation from brain mri images using deep learning,'' in \emph{2017 IEEE 14th International Symposium on Biomedical Imaging (ISBI 2017)}.\hskip 1em plus 0.5em minus 0.4em\relax IEEE, 2017, pp. 849--852.

\bibitem{ueda2019age}
M.~Ueda, K.~Ito, K.~Wu, K.~Sato, Y.~Taki, H.~Fukuda, and T.~Aoki, ``An age estimation method using 3d-cnn from brain mri images,'' in \emph{2019 IEEE 16th international symposium on biomedical imaging (ISBI 2019)}.\hskip 1em plus 0.5em minus 0.4em\relax IEEE, 2019, pp. 380--383.

\bibitem{jonsson2019brain}
B.~A. J{\'o}nsson, G.~Bjornsdottir, T.~E. Thorgeirsson, L.~M. Ellingsen, G.~B. Walters, D.~F. Gudbjartsson, H.~Stefansson, K.~Stefansson, and M.~O. Ulfarsson, ``Brain age prediction using deep learning uncovers associated sequence variants,'' \emph{Nature communications}, vol.~10, no.~1, p. 5409, 2019.

\bibitem{feng2020estimating}
X.~Feng, Z.~C. Lipton, J.~Yang, S.~A. Small, F.~A. Provenzano, A.~D.~N. Initiative, F.~L. D.~N. Initiative \emph{et~al.}, ``Estimating brain age based on a uniform healthy population with deep learning and structural magnetic resonance imaging,'' \emph{Neurobiology of aging}, vol.~91, pp. 15--25, 2020.

\bibitem{bashyam2020mri}
V.~M. Bashyam, G.~Erus, J.~Doshi, M.~Habes, I.~M. Nasrallah, M.~Truelove-Hill, D.~Srinivasan, L.~Mamourian, R.~Pomponio, Y.~Fan \emph{et~al.}, ``Mri signatures of brain age and disease over the lifespan based on a deep brain network and 14 468 individuals worldwide,'' \emph{Brain}, vol. 143, no.~7, pp. 2312--2324, 2020.

\bibitem{fawaz2021benchmarking}
A.~Fawaz, L.~Z. Williams, A.~Alansary, C.~Bass, K.~Gopinath, M.~da~Silva, S.~Dahan, C.~Adamson, B.~Alexander, D.~Thompson \emph{et~al.}, ``Benchmarking geometric deep learning for cortical segmentation and neurodevelopmental phenotype prediction,'' \emph{bioRxiv}, pp. 2021--12, 2021.

\bibitem{vosylius2020geometric}
V.~Vosylius, A.~Wang, C.~Waters, A.~Zakharov, F.~Ward, L.~Le~Folgoc, J.~Cupitt, A.~Makropoulos, A.~Schuh, D.~Rueckert \emph{et~al.}, ``Geometric deep learning for post-menstrual age prediction based on the neonatal white matter cortical surface,'' in \emph{International Workshop on Uncertainty for Safe Utilization of Machine Learning in Medical Imaging}.\hskip 1em plus 0.5em minus 0.4em\relax Springer, 2020, pp. 174--186.

\bibitem{monti2017geometric}
F.~Monti, D.~Boscaini, J.~Masci, E.~Rodola, J.~Svoboda, and M.~M. Bronstein, ``Geometric deep learning on graphs and manifolds using mixture model cnns,'' in \emph{Proceedings of the IEEE conference on computer vision and pattern recognition}, 2017, pp. 5115--5124.

\bibitem{cohen2018spherical}
T.~S. Cohen, M.~Geiger, J.~K{\"o}hler, and M.~Welling, ``Spherical cnns,'' \emph{arXiv preprint arXiv:1801.10130}, 2018.

\bibitem{defferrard2016convolutional}
M.~Defferrard, X.~Bresson, and P.~Vandergheynst, ``Convolutional neural networks on graphs with fast localized spectral filtering,'' \emph{Advances in neural information processing systems}, vol.~29, 2016.

\bibitem{kipf2016semi}
T.~Kipf, ``Semi-supervised classification with graph convolutional networks,'' \emph{arXiv preprint arXiv:1609.02907}, 2016.

\bibitem{zhao2019spherical}
F.~Zhao, S.~Xia, Z.~Wu, D.~Duan, L.~Wang, W.~Lin, J.~H. Gilmore, D.~Shen, and G.~Li, ``Spherical u-net on cortical surfaces: methods and applications,'' in \emph{International Conference on Information Processing in Medical Imaging}.\hskip 1em plus 0.5em minus 0.4em\relax Springer, 2019, pp. 855--866.

\bibitem{qi2017pointnet++}
C.~R. Qi, L.~Yi, H.~Su, and L.~J. Guibas, ``Pointnet++: Deep hierarchical feature learning on point sets in a metric space,'' \emph{Advances in neural information processing systems}, vol.~30, 2017.

\bibitem{zhao2024transformer}
H.~Zhao, H.~Cai, and M.~Liu, ``Transformer based multi-modal mri fusion for prediction of post-menstrual age and neonatal brain development analysis,'' \emph{Medical Image Analysis}, vol.~94, p. 103140, 2024.

\bibitem{dahan2024multiscale}
S.~Dahan, L.~Z. Williams, D.~Rueckert, and E.~C. Robinson, ``The multiscale surface vision transformer,'' \emph{ArXiv}, pp. arXiv--2303, 2024.

\bibitem{gu2024mamba}
A.~Gu and T.~Dao, ``Mamba: Linear-time sequence modeling with selective state spaces,'' in \emph{First conference on language modeling}, 2024.

\bibitem{cherubini2016importance}
A.~Cherubini, M.~E. Caligiuri, P.~P{\'e}ran, U.~Sabatini, C.~Cosentino, and F.~Amato, ``Importance of multimodal mri in characterizing brain tissue and its potential application for individual age prediction,'' \emph{IEEE journal of biomedical and health informatics}, vol.~20, no.~5, pp. 1232--1239, 2016.

\bibitem{kaufmann2019common}
T.~Kaufmann, D.~van~der Meer, N.~T. Doan, E.~Schwarz, M.~J. Lund, I.~Agartz, D.~Aln{\ae}s, D.~M. Barch, R.~Baur-Streubel, A.~Bertolino \emph{et~al.}, ``Common brain disorders are associated with heritable patterns of apparent aging of the brain,'' \emph{Nature neuroscience}, vol.~22, no.~10, pp. 1617--1623, 2019.

\bibitem{hof1996neuropathological}
P.~Hof, P.~Giannakopoulos, and C.~Bouras, ``The neuropathological changes associated with normal brain aging,'' \emph{Histology and histopathology}, 1996.

\bibitem{raz2010trajectories}
N.~Raz, P.~Ghisletta, K.~M. Rodrigue, K.~M. Kennedy, and U.~Lindenberger, ``Trajectories of brain aging in middle-aged and older adults: regional and individual differences,'' \emph{Neuroimage}, vol.~51, no.~2, pp. 501--511, 2010.

\bibitem{popescu2021local}
S.~G. Popescu, B.~Glocker, D.~J. Sharp, and J.~H. Cole, ``Local brain-age: a u-net model,'' \emph{Frontiers in Aging Neuroscience}, vol.~13, p. 761954, 2021.

\bibitem{gianchandani2023voxel}
N.~Gianchandani, M.~Dibaji, J.~Ospel, F.~Vega, M.~Bento, M.~E. MacDonald, and R.~Souza, ``A voxel-level approach to brain age prediction: A method to assess regional brain aging,'' \emph{arXiv preprint arXiv:2310.11385}, 2023.

\bibitem{makropoulos2018developing}
A.~Makropoulos, E.~C. Robinson, A.~Schuh, R.~Wright, S.~Fitzgibbon, J.~Bozek, S.~J. Counsell, J.~Steinweg, K.~Vecchiato, J.~Passerat-Palmbach \emph{et~al.}, ``The developing human connectome project: A minimal processing pipeline for neonatal cortical surface reconstruction,'' \emph{Neuroimage}, vol. 173, pp. 88--112, 2018.

\bibitem{makropoulos2014automatic}
A.~Makropoulos, I.~S. Gousias, C.~Ledig, P.~Aljabar, A.~Serag, J.~V. Hajnal, A.~D. Edwards, S.~J. Counsell, and D.~Rueckert, ``Automatic whole brain mri segmentation of the developing neonatal brain,'' \emph{IEEE transactions on medical imaging}, vol.~33, no.~9, pp. 1818--1831, 2014.

\bibitem{schuh2017deformable}
A.~Schuh, A.~Makropoulos, R.~Wright, E.~C. Robinson, N.~Tusor, J.~Steinweg, E.~Hughes, L.~C. Grande, A.~Price, J.~Hutter \emph{et~al.}, ``A deformable model for the reconstruction of the neonatal cortex,'' in \emph{2017 IEEE 14th international symposium on biomedical imaging (ISBI 2017)}.\hskip 1em plus 0.5em minus 0.4em\relax IEEE, 2017, pp. 800--803.

\bibitem{kendall2018multi}
A.~Kendall, Y.~Gal, and R.~Cipolla, ``Multi-task learning using uncertainty to weigh losses for scene geometry and semantics,'' in \emph{Proceedings of the IEEE conference on computer vision and pattern recognition}, 2018, pp. 7482--7491.

\bibitem{wang2021pyramid}
W.~Wang, E.~Xie, X.~Li, D.-P. Fan, K.~Song, D.~Liang, T.~Lu, P.~Luo, and L.~Shao, ``Pyramid vision transformer: A versatile backbone for dense prediction without convolutions,'' in \emph{Proceedings of the IEEE/CVF international conference on computer vision}, 2021, pp. 568--578.

\bibitem{wang2022pvt}
------, ``Pvt v2: Improved baselines with pyramid vision transformer,'' \emph{Computational visual media}, vol.~8, no.~3, pp. 415--424, 2022.

\bibitem{yu2022metaformer}
W.~Yu, M.~Luo, P.~Zhou, C.~Si, Y.~Zhou, X.~Wang, J.~Feng, and S.~Yan, ``Metaformer is actually what you need for vision,'' in \emph{Proceedings of the IEEE/CVF conference on computer vision and pattern recognition}, 2022, pp. 10\,819--10\,829.

\bibitem{otmakhov1993measuring}
N.~Otmakhov, A.~M. Shirke, and R.~Malinow, ``Measuring the impact of probabilistic transmission on neuronal output,'' \emph{Neuron}, vol.~10, no.~6, pp. 1101--1111, 1993.

\bibitem{aitchison2016hamiltonian}
L.~Aitchison and M.~Lengyel, ``The hamiltonian brain: Efficient probabilistic inference with excitatory-inhibitory neural circuit dynamics,'' \emph{PLoS computational biology}, vol.~12, no.~12, p. e1005186, 2016.

\bibitem{kelly2024cortical}
C.~E. Kelly, D.~K. Thompson, C.~L. Adamson, G.~Ball, T.~Dhollander, R.~Beare, L.~G. Matthews, B.~Alexander, J.~L. Cheong, L.~W. Doyle \emph{et~al.}, ``Cortical growth from infancy to adolescence in preterm and term-born children,'' \emph{Brain}, vol. 147, no.~4, pp. 1526--1538, 2024.

\bibitem{bouyssi2018altered}
M.~Bouyssi-Kobar, J.~Murnick, M.~Brossard-Racine, T.~Chang, E.~Mahdi, M.~Jacobs, and C.~Limperopoulos, ``Altered cerebral perfusion in infants born preterm compared with infants born full term,'' \emph{The Journal of Pediatrics}, vol. 193, pp. 54--61, 2018.

\bibitem{ball2013development}
G.~Ball, L.~Srinivasan, P.~Aljabar, S.~J. Counsell, G.~Durighel, J.~V. Hajnal, M.~A. Rutherford, and A.~D. Edwards, ``Development of cortical microstructure in the preterm human brain,'' \emph{Proceedings of the National Academy of Sciences}, vol. 110, no.~23, pp. 9541--9546, 2013.

\bibitem{zheng2021developmental}
W.~Zheng, Z.~Zhao, Z.~Zhang, T.~Liu, Y.~Zhang, J.~Fan, and D.~Wu, ``Developmental pattern of the cortical topology in high-functioning individuals with autism spectrum disorder,'' \emph{Human Brain Mapping}, vol.~42, no.~3, pp. 660--675, 2021.

\bibitem{seidlitz2018morphometric}
J.~Seidlitz, F.~V{\'a}{\v{s}}a, M.~Shinn, R.~Romero-Garcia, K.~J. Whitaker, P.~E. V{\'e}rtes, K.~Wagstyl, P.~K. Reardon, L.~Clasen, S.~Liu \emph{et~al.}, ``Morphometric similarity networks detect microscale cortical organization and predict inter-individual cognitive variation,'' \emph{Neuron}, vol.~97, no.~1, pp. 231--247, 2018.

\bibitem{liu2021diffusion}
T.~Liu, F.~Gao, W.~Zheng, Y.~You, Z.~Zhao, Y.~Lv, W.~Chen, H.~Zhang, C.~Ji, and D.~Wu, ``Diffusion mri of the infant brain reveals unique asymmetry patterns during the first-half-year of development,'' \emph{NeuroImage}, vol. 242, p. 118465, 2021.

\bibitem{duboc2015asymmetry}
V.~Duboc, P.~Dufourcq, P.~Blader, and M.~Roussign{\'e}, ``Asymmetry of the brain: development and implications,'' \emph{Annual review of genetics}, vol.~49, no.~1, pp. 647--672, 2015.

\bibitem{kong2018mapping}
X.-Z. Kong, S.~R. Mathias, T.~Guadalupe, E.~L.~W. Group, D.~C. Glahn, B.~Franke, F.~Crivello, N.~Tzourio-Mazoyer, S.~E. Fisher, P.~M. Thompson \emph{et~al.}, ``Mapping cortical brain asymmetry in 17,141 healthy individuals worldwide via the enigma consortium,'' \emph{Proceedings of the National Academy of Sciences}, vol. 115, no.~22, pp. E5154--E5163, 2018.

\bibitem{nam2015alterations}
K.~W. Nam, N.~Castellanos, A.~Simmons, S.~Froudist-Walsh, M.~P. Allin, M.~Walshe, R.~M. Murray, A.~Evans, J.-S. Muehlboeck, and C.~Nosarti, ``Alterations in cortical thickness development in preterm-born individuals: Implications for high-order cognitive functions,'' \emph{NeuroImage}, vol. 115, pp. 64--75, 2015.

\bibitem{mento2010functional}
G.~Mento, A.~Suppiej, G.~Alto{\`e}, and P.~S. Bisiacchi, ``Functional hemispheric asymmetries in humans: electrophysiological evidence from preterm infants,'' \emph{European Journal of Neuroscience}, vol.~31, no.~3, pp. 565--574, 2010.

\bibitem{schmitz2023altered}
B.~Schmitz-Koep, A.~Menegaux, C.~Gaser, E.~Brandes, D.~Schinz, M.~Thalhammer, M.~Daamen, H.~Boecker, C.~Zimmer, J.~Priller \emph{et~al.}, ``Altered gray matter cortical and subcortical t1-weighted/t2-weighted ratio in premature-born adults,'' \emph{Biological Psychiatry: Cognitive Neuroscience and Neuroimaging}, vol.~8, no.~5, pp. 495--504, 2023.

\bibitem{zhang2025dynamic}
Z.~Zhang, C.~Zhang, X.~Zhang, Y.~Xu, W.~Dou, M.~Li, W.~Zheng, and B.~Li, ``Dynamic structure-function coupling in macroscale neonatal brain networks,'' \emph{Communications Biology}, 2025.

\bibitem{dimitrova2020heterogeneity}
R.~Dimitrova, M.~Pietsch, D.~Christiaens, J.~Ciarrusta, T.~Wolfers, D.~Batalle, E.~Hughes, J.~Hutter, L.~Cordero-Grande, A.~N. Price \emph{et~al.}, ``Heterogeneity in brain microstructural development following preterm birth,'' \emph{Cerebral Cortex}, vol.~30, no.~9, pp. 4800--4810, 2020.

\bibitem{nivins2025gestational}
S.~Nivins, N.~Padilla, H.~Kvanta, and U.~{\AA}d{\'e}n, ``Gestational age and cognitive development in childhood,'' \emph{JAMA Network Open}, vol.~8, no.~4, pp. e254\,580--e254\,580, 2025.

\bibitem{kline2020early}
J.~E. Kline, V.~S.~P. Illapani, L.~He, M.~Altaye, J.~W. Logan, and N.~A. Parikh, ``Early cortical maturation predicts neurodevelopment in very preterm infants,'' \emph{Archives of Disease in Childhood-Fetal and Neonatal Edition}, vol. 105, no.~5, pp. 460--465, 2020.

\bibitem{hinojosa2017clinical}
M.~Hinojosa-Rodr{\'\i}guez, T.~Harmony, C.~Carrillo-Prado, J.~D. Van~Horn, A.~Irimia, C.~Torgerson, and Z.~Jacokes, ``Clinical neuroimaging in the preterm infant: diagnosis and prognosis,'' \emph{NeuroImage: Clinical}, vol.~16, pp. 355--368, 2017.

\end{thebibliography}

\end{document}